\documentclass[12pt, a4paper]{article}
\pdfoutput=1

\usepackage{amsmath}
\usepackage{amsfonts}
\usepackage{amssymb}
\usepackage{graphicx, rotating}
\usepackage{epstopdf}

\usepackage{epsfig}
\usepackage{latexsym}
\usepackage{graphicx}
\usepackage{color}
\usepackage{amsmath,amssymb}
\usepackage{cite}
\usepackage{slashed}

\usepackage{hyperref}
\hypersetup{colorlinks, citecolor=bluscuro, linkcolor=black, urlcolor=bluscuro}
\definecolor{rossos}{cmyk}{0,1,1,0.55}
\definecolor{bluscuro}{rgb}{0.15, 0.2, .85}
\definecolor{bluchiaro}{cmyk}{1,.3,0.,0.1}

\numberwithin{equation}{section}

\makeatletter
\def\section{\@startsection {section}{1}{\z@}{-3.5ex plus -1ex minus
-.2ex}{2.3ex plus .2ex}{\large\bf}}
\def\subsection{\@startsection{subsection}{2}{\z@}{-3.25ex plus -1ex
minus -.2ex}{1.5ex plus .2ex}{\normalsize\bf}}
\makeatother
\newcommand{\captionfonts}{\small}
\makeatletter  
\long\def\@makecaption#1#2{%
 \vskip\abovecaptionskip
 \sbox\@tempboxa{{\captionfonts #1: #2}}%
 \ifdim \wd\@tempboxa >\hsize
   {\captionfonts #1: #2\par}
 \else
   \hbox to\hsize{\hfil\box\@tempboxa\hfil}%
 \fi
 \vskip\belowcaptionskip}
\makeatother   

\catcode`@=11
\def\marginnote#1{}
\newcount\hour
\newcount\minute
\newtoks\amorpm
\hour=\time\divide\hour
by60
\minute=\time{\multiply\hour by60 \global\advance\minute
by-\hour}
\edef\standardtime{{\ifnum\hour<12 \global\amorpm={am}
\else\global\amorpm={pm}\advance\hour by-12 \fi
\ifnum\hour=0
\hour=12 \fi
\number\hour:\ifnum\minute<10
0\fi\number\minute\the\amorpm}}
\edef\militarytime{\number\hour:\ifnum\minute<10
0\fi\number\minute}
\def\draftlabel#1{{\@bsphack\if@filesw
{\let\thepage\relax
\xdef\@gtempa{\write\@auxout{\string
\newlabel{#1}{{\@currentlabel}{\thepage}}}}}\@gtempa
\if@nobreak
\ifvmode\nobreak\fi\fi\fi\@esphack}
\gdef\@eqnlabel{#1}}
\def\@eqnlabel{}
\def\@vacuum{}
\def\draftmarginnote#1{\marginpar{\raggedright\scriptsize\tt#1}}
\def\draft{\oddsidemargin
0.0truein
\def\@oddfoot{\sl preliminary draft \hfil
\rm\thepage\hfil\sl\today\quad\militarytime}
\let\@evenfoot\@oddfoot
\overfullrule 3pt
\let\label=\draftlabel
\let\marginnote=\draftmarginnote
\def\@eqnnum{(\theequation)\rlap{\kern\marginparsep\tt\@eqnlabel}
\global\let\@eqnlabel\@vacuum}
}
\catcode`@=12
\def\dj{\hbox{d\kern-0.347em \vrule width 0.3em height 1.252ex depth
-1.21ex \kern 0.051em}}

\def\ee{{\rm e}\,}

\def\ba{\bar a}

\def\Dirac{\,\raise.15ex\hbox{/}\mkern-13.5mu D}
\def\dirac{\,\raise.15ex\hbox{/}\kern-.57em \partial}
\def\aslash{\,\raise.15ex\hbox{/}\mkern-13.5mu A}
\def\shalf{{\ifinner {\textstyle \frac{1}{2}}\else \frac{1}{2} \fi}}
\def\sthreehalfs{{\ifinner {\textstyle \frac{3}{2}}\else \frac{3}{2} \fi}}
\def\sshalf{{\ifinner {\scriptstyle \frac{1}{2}}\else \frac{1}{2} \fi}}
\def\sfourth{{\ifinner {\textstyle \frac{1}{4}}\else frac{1}{4} \fi}}
\def\sphifour{{\ifinner {\textstyle \frac{1}{4!}}\else \frac{1}{4!} \fi}}

\def\XXint#1#2#3{{\setbox0=\hbox{$#1{#2#3}{\int}$}
    \vcenter{\hbox{$#2#3$}}\kern-.5\wd0}}

\def\bea{\begin{eqnarray}} \def\eea{\end{eqnarray}}
\def\be{\begin{eqnarray}} \def\ee{\end{eqnarray}} \def\nn{\nonumber}

\newcommand{\promille}{%
 \relax\ifmmode\promillezeichen
       \else\leavevmode\(\mathsurround=0pt\promillezeichen\)\fi}
\newcommand{\promillezeichen}{%
 \kern-.05em%
 \raise.5ex\hbox{\the\scriptfont0 0}%
 \kern-.15em/\kern-.15em%
 \lower.25ex\hbox{\the\scriptfont0 00}}
\newcommand{\beq}{\begin{eqnarray}}
\newcommand{\eeq}{\end{eqnarray}}

\newcommand{\desl}{\partial\hspace{-4.9pt}{\scriptstyle /}}

\def\nn{\nonumber}

\newcommand{\GeV}{{\rm \,GeV}}
\newcommand{\TeV}{{\rm \,TeV}}

\def\cs2{c_{s}^{2}}

 \def\be   {\begin{equation}}   \def\ee   {\end{equation}}
 \def\ba   {\begin{array}}      \def\ea   {\end{array}}
 \def\bea  {\begin{eqnarray}}   \def\eea  {\end{eqnarray}}
 \def\bean {\begin{eqnarray*}}  \def\eean {\end{eqnarray*}}

\def\hhref#1{\href{http://arxiv.org/abs/#1}{#1}} 

\begin{document}

{\footnotesize  SACLAY-T11/031 \hfill CERN-PH-TH/2011-040}
\vspace{5mm}
\vspace{0.5cm}
\begin{center}

\def\thefootnote{\fnsymbol{footnote}}

{\Large \bf On the Importance of Electroweak Corrections\\
\vskip 0.3cm
for Majorana Dark Matter Indirect Detection
}
\\[1.2cm]
{\large Paolo Ciafaloni $^{\rm a}$, Marco Cirelli $^{\rm b,c}$, Denis Comelli $^{\rm d}$} \\[0.2cm]
{\large Andrea De Simone $^{\rm e}$, Antonio Riotto $^{\rm b,f}$, Alfredo Urbano $^{\rm a,g}$}
\\[1cm]

{\small \textit{$^{\rm a}$ Dipartimento di Fisica, Universit\`a di Lecce and INFN - Sezione di
Lecce, \\Via per Arnesano, I-73100 Lecce, Italy }}

{\small \textit{$^{\rm b}$  CERN, PH-TH Division, CH-1211,
Gen\`eve 23,  Switzerland}}

{\small \textit{$^{\rm c}$ Institut de Physique Th\'eorique, CNRS URA 2306\\ and CEA/Saclay, F-91191
 Gif-sur-Yvette, France }}

{\small \textit{$^{\rm d}$  INFN - Sezione di Ferrara, Via Saragat 3, I-44100 Ferrara, Italy}}

{\small \textit{$^{\rm e}$ Institut de Th\'eorie des Ph\'enom\`enes Physiques,\\
 \'Ecole Polytechnique F\'ed\'erale de Lausanne, CH-1015 Lausanne,
Switzerland}}

{\small \textit{$^{\rm f}$  INFN, Sezione di Padova, Via Marzolo 8, I-35131, Padova, Italy}}

{\small \textit{$^{\rm g}$  IFAE, Universitat Aut\`onoma de Barcelona, 08193 Bellaterra, Barcelona, Spain}}

\vspace{.2cm}

\end{center}

\vspace{.8cm}

\begin{center}
\textbf{Abstract}
\end{center}

\noindent
Recent analyses have shown that the inclusion of electroweak corrections can alter significantly
 the energy spectra of Standard Model particles originated from  dark matter annihilations.
We investigate the important situation where the radiation of electroweak
gauge bosons has a substantial influence: a Majorana dark matter particle annihilating into two 
light fermions.
This process is in $p$-wave and hence suppressed by the small value of the
relative velocity of the annihilating particles.
The  inclusion of electroweak radiation eludes this suppression and 
 opens up a potentially sizeable $s$-wave contribution to the annihilation cross section.
We study this effect in detail and explore its impact on the fluxes of stable particles resulting
from the dark matter annihilations, which are relevant
for dark matter indirect searches.
We also discuss the effective field theory approach, pointing out that the opening of the $s$-wave 
is missed at the level of dimension-six operators and only encoded by higher orders.

\def\thefootnote{\arabic{footnote}}
\setcounter{footnote}{0}
\pagestyle{empty}

\newpage
\pagestyle{plain}
\setcounter{page}{1}

\section{Introduction}
\noindent

\noindent
In a recent paper \cite{paper1} some of us have  pointed out that
 the energy spectra of the Standard Model (SM) particles originating from Dark Matter
 (DM) annihilation/decays can be significantly affected  by  ElectroWeak (EW)  corrections,
if the mass $M$ of the DM particles is larger than the EW scale (here set to be the mass of the $W$ boson
$m_W$).
The emission of EW gauge bosons from the final state of the annihilation/decay process is enhanced by single logarithms $\ln M^2/m_W^2$ in the collinear region and by double logarithms $\ln^2 M^2/m_W^2$ when both collinear and infrared singularities are present, and
implies  that all stable particles of the SM appear in the final
spectrum, independently of the primary annihilation/decay channel.
The inclusion of EW corrections seems  therefore an essential ingredient   in indirect
searches for DM.
The impact of EW corrections is particularly relevant in two situations: (1) when one is interested in
the energy region of the  final fluxes
(after propagation from the source) which corresponds to the low-energy tail of the spectrum,
populated by the decay products of the additional gauge bosons; (2) when some of the stable
 species appear only if EW corrections are taken into account, for instance antiprotons (from $W$ or $Z$ decays)
 in an otherwise purely leptonic annihilation.

One basic assumption made in  Ref. \cite{paper1}  was that the tree-level $2\rightarrow 2$ annihilation cross section   of the DM particles into SM states was dominant over the $2\rightarrow 3$ cross section with soft gauge boson emission from the external legs,
and the latter was factorized with respect to the former.

While this assumption is certainly reasonable and commonly made (for instance if the DM is a heavy Dirac fermion singlet
under the SM gauge group), there are well-motivated cases in which it is questionable.
Consider for instance a DM particle $\chi$ which is a Majorana fermion and a SM singlet.
The  cross section of $\chi\chi\to f\bar f$, where DM annihilates into SM  fermions of mass $m_f$,  consists of a velocity-independent
($s$-wave) and a velocity-dependent ($p$-wave) contribution
\be
v\sigma=a+b\,v^2+\mathcal{O}(v^4)\,,
\ee
where $v\sim 10^{-3}$ is the relative velocity (in units of $c$) of the DM states in our Galaxy.
By helicity arguments, $a\propto (m_f/M)^2$ and hence very suppressed for
light final state fermions (e.g.~leptons), while the $p$-wave term is suppressed by $v^2$.
In this case, it is not  guaranteed that the 2-body
annihilation cross section is quantitatively larger than the one with EW corrections.
Indeed, the latter ones may open a sizeable  $s$-wave contribution and elude the suppressions.

The scope of this paper is therefore to generalize the results
of Ref. \cite{paper1} to the interesting case in which the  2-body annihilation cross-section is not automatically
larger than the one with soft gauge boson emission. The same kind of
effect has been considered in the past with respect to photon radiation in Refs.~\cite{bergstrom1,bergstrom2,bergstrom3};
this work is partly, but not only,  an extension of those analyses to include also $W, Z$
 gauge bosons. On the other hand, an approach similar to ours is the one carried out
in Ref. \cite{w}, but our final results disagree with the ones published there
\footnote{The disagreement originates from the use of incorrect Fierz identities, thus invalidating
the calculation of the cross sections,
as we have been informed by the authors of Ref.~\cite{w} in a private communication.}.
Other works which considered at various levels the impact of EW corrections on DM 
annihilation or cosmic ray physics include~\cite{list1}.
For somewhat related work on 3-body annihilations below threshold see also Ref.~\cite{list2}.

We shall show in detail that, whenever the dark matter annihilation occurs by exchange of a heavy 
intermediate state, at the lowest order in the expansion in inverse powers of this heavy mass,  
final state radiation is not sufficient to remove
the helicity suppression, while it is  efficiently removed at  higher orders by 
processes involving the emission both from external legs  and from virtual internal propagators.
Although subleading in terms of powers of the heavy mass, these contributions  do not pay the
velocity suppression and actually can be dominant.

This allows us to raise an important and cautionary remark concerning the use of the
effective field theory approach to describe DM interactions \cite{eff}.
If the interactions of DM with SM fermions are described  by
effective four-fermion dimension-six operators,
then the emission of soft gauge bosons can only take place at the lowest order from
the external legs and the corresponding cross section remains helicity suppressed.
In other words, the effect of opening up a large $s$-wave annihilation channel is missed at the level of dimension-six operators
and then one may (incorrectly) conclude that the whole cross section is still suppressed.
Instead, as we shall point out, the diagrams leading to $s$-wave contributions correspond to  operators with
dimension higher than six, whose quantitative relevance for the cross section can be comparable or larger than that due to dimension-six operators,
despite the larger dimensionality.

The plan of the paper is as follows.
In  Section \ref{sec:generalsettings}  we describe  the simple model we shall use throughout the paper.
Then, Section \ref{sec:2body} introduces preliminary considerations about 2-body annihilations,
especially about helicity-suppression, setting the ground for the subsequent discussion.
 Section \ref{sec:3body}  contains the calculations and the results for the annihilation cross section with the inclusion of
 EW bremsstrahlung (including the Ward Identities  check and the remarks on the effective field theory approach).
Our results for the 3-body cross sections are in disagreement
with those of Ref.  \cite{w}, while we find a perfect agreement with those in Ref.  \cite{bergstrom1} concerning 
the radiation of one photon.
With these analytical results at hand, we are then ready to study in detail the impact
of the opening of the $s$-wave on the fluxes of stable SM particles. To this end, we carry out a numerical
analysis. In Section \ref{sec:spectra}, we derive the energy spectra at production while in Section
 \ref{sec:propagation} these spectra are then propagated to give the fluxes of particles at detection.
Concluding remarks are collected in Section \ref{sec:conclusions}.

\section{The model}
\label{sec:generalsettings}

In this section we present the (toy) model we shall use in the paper to describe the relevance
of the EW corrections in DM annihilations.
Let us add to the particle content of the SM a
 Majorana spinor $\chi$ with mass $M_\chi$,  singlet under the SM gauge
 group and playing the role of DM,
and a scalar $SU(2)$-doublet $S$, with mass $M_S>M_\chi$
\begin{equation}
\chi=\chi^C\qquad
S=\left(%
\begin{array}{c}
\eta^+ \\
\eta^0%
\end{array}%
\right)\,.
\end{equation}
The field $S$ provides the interactions
of the DM with the generic fermion of the SM, described by the left-handed doublet
 $L=(f_1, f_2)$. In fact, the total  Lagrangian of the model is (see also Ref.~\cite{ma})
\begin{equation}
{\cal L}={\cal L}_{\rm SM}+{\cal L}_\chi+{\cal L}_S+{\cal L}_{\rm int}\,,
\label{eq:Lagrangian}
\end{equation}
where to the Standard Model Lagrangian $\cal{L}_{\rm SM}$ we added
\bea
{\cal L}_\chi&=&{1\over 2}{\bar{\chi}(i\desl-M_\chi)\chi}\,,\\
{\cal L}_S&=&(D_\mu S)^\dagger (D^\mu S)-M_S^2S^\dagger S \,, \\
{\cal L}_{\rm int}&=&y_L \bar{\chi}(Li\sigma_2 S)+{\rm h.c.}
=y_L(\bar{\chi}P_L f_2 \eta^+-\bar{\chi}P_Lf_1\eta^0)+{\rm h.c.} \,,
\label{eq:Lagrangianint}
\eea
where the 4-component notation has been used and where
contractions on $SU(2)$ indices is defined as  
$(Li\sigma_2 S)\equiv L_i(i\sigma_2)_{ij}S_j$.
Moreover, we shall adopt the convention for projectors: $P_{R, L}=(1\pm \gamma^5)/2$.
The stability of the DM can be achieved e.g. by endowing $\chi$ and $S$ with odd parity
under a $Z_2$ symmetry, while the rest of the SM spectrum is even.

The  model is manifestly gauge invariant, and is the same of Ref. \cite{w}, which will allow us a direct comparison between their results and ours.
A reader expert in supersymmetry would recognize the same interactions of a Bino
with fermions and their supersymmetric scalar partners.

We shall restrict our attention to the
massless limit $m_{f_1}=m_{f_2}=0$. While reasonable for leptons and light quarks, this approximation may not be good for heavy quarks.
For instance, the DM annihilation into $t\bar t$, if kinematically allowed, would proceed through $s$-wave with a contribution proportional to $(m_t/M_\chi)^2$,
which can be large
already without  EW corrections.
However, the generalization of our calculations to non-zero fermion masses is beyond the scope
of this paper. 

As anticipated in the Introduction, one of the main goals of the present paper is to show
that the inclusion of higher-order processes with emission of soft weak gauge bosons
 evades the helicity suppression and turns on an unsuppressed $s$-wave contribution to
 the DM annihilation cross section.
Before turning to the details of the calculation of the $2\to 3$ scatterings in Section \ref{sec:3body}, let us first review some standard material about $2\to 2$ annihilations, with particular emphasis
on the role of helicity suppression. This will  serve to set the notation
and to highlight the main points for later use. The analogous results for the amplitudes and the cross sections of
the 2-body and 3-body processes in the case of Dirac DM are reported
in Appendix \ref{app:dirac}.

\section{Two-body annihilation into fermions and the helicity suppression}
\label{sec:2body}

\begin{figure}[t]
\begin{center}
 \includegraphics[width=12 cm]{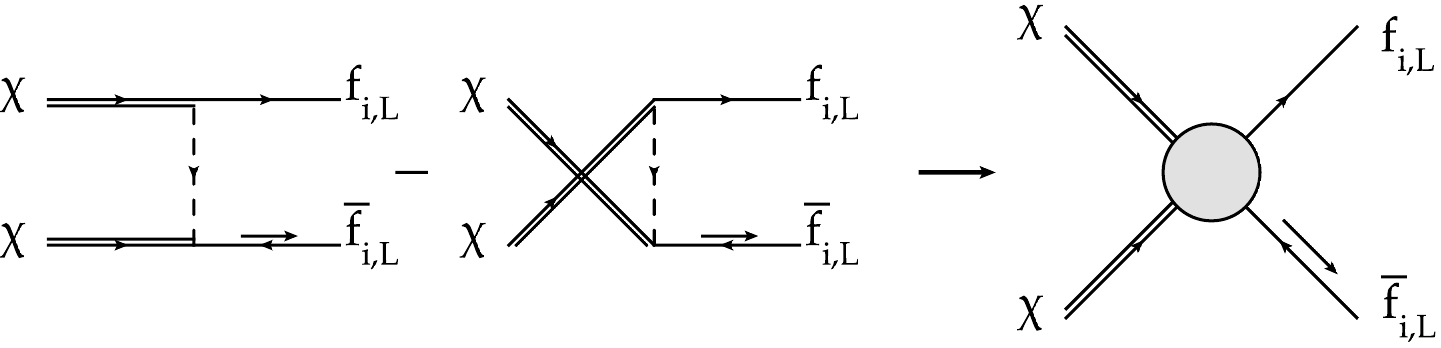}
 \caption{\emph{Feynman diagrams for the tree-level annihilation in Eq.~(\ref{eq:2BodyBinoAnnihilation}) together with its effective contraction in the limit $M_{S}\gg M_{\chi}$.}}
 \label{fig:2BodyTreeLevel}
 \end{center}
\end{figure}

\noindent
Let us consider the annihilation of the DM Majorana fermion into a pair of massless left-handed fermions
(see Figure~\ref{fig:2BodyTreeLevel})
\begin{equation}
\label{eq:2BodyBinoAnnihilation}
\chi(k_1)
\chi(k_2)\to f_{L_i}(p_1)\bar f_{L_i}(p_2)\, .
\end{equation}
The cross section admits the usual expansion in powers of  the relative velocity  $v$
 of the initial DM particles
\begin{equation}
\label{eq:CrossSectionExpansion}
 v\sigma =a+b\,v^{2}+\mathcal{O}(v^4)\ ,
\end{equation}
where
the coefficients $a, b$ corresponding to $s$- and $p$-waves, respectively,
are given by
\be
a=0\, ,\qquad b=\frac{|y_L|^4}{48\pi}\frac{1+r^2}{(1+r)^4}{1\over M_{\chi}^2}\, ,
\label{bpwave}
\ee
where we have defined 
\be
r\equiv {M_{S}^2\over M_{\chi}^2}\,.
\ee
This result shows the well-known fact that the first non-zero contribution to the tree-level cross section for the Majorana DM is
velocity dependent, and hence suppressed.

Let us try to understand  this fact in simple terms.
At the level of Feynman diagrams the  Majorana nature of the  DM  implies the
presence of two crossed channels,  $t$ and $u$ (for  Dirac DM,   only the $t$-channel contribution would
be present).
The amplitudes are of the form $(\bar \chi P_Lf)(\bar{f}P_R\chi)$, which becomes $(\bar \chi \gamma_\alpha P_R \chi )(\bar{f}\gamma^{\alpha}P_L f)$ after chiral Fierz transformation.
The total tree-level amplitude for the process in Eq.~(\ref{eq:2BodyBinoAnnihilation}) is given by
\be
\label{currentX}
{\cal M}_{0}= \frac{i|y_L|^2}{2}
[\bar u_f(p_1)\;\gamma_\alpha P_L v_f(p_2)] \left[\frac{D_{11}-D_{12}}{2}  \bar v_{\chi}(k_2)\gamma^{\alpha}  u_{\chi}(k_1)+\frac{D_{11}+ D_{12}}{2} \bar v_{\chi}(k_2)\gamma^{\alpha}\gamma_5 u_{\chi}(k_1)
\right]\, ,
\ee
where we have defined the quantities
\be\label{dd}
D_{ij}\equiv \frac{1}{(p_i-k_j)^2-r M_\chi^2}\, ,
\ee
which satisfy the property $D_{i1}-D_{i2} =2\,p_i \cdot (k_1-k_2)D_{i1}D_{i2}$.
Notice that the momenta of the incoming DM particles are such that
$k_1^\mu-k_2^\mu\sim \mathcal{O}(v)M_\chi$. We thus obtain that
\be
\label{di1di2}
D_{i1}-D_{i2}
\sim  {\cal O}(v)M_{\chi}^2\; D_{i1}\,D_{i 2}\, .
\ee
In the matrix element in Eq.~(\ref{currentX}), the first term represents a  vector current while the
second term is an  axial-vector current. Let us  analyze the velocity factors present in each of them,
in the non-relativistic limit $v\ll 1$.
The vector current  is multiplied by a factor proportional to $v$ due to Eq.~(\ref{di1di2}).
 For the axial current of Eq.~(\ref{currentX}), using
the Gordon identities we have
\be
\label{GordonId}
\bar{v}_{\chi}(k_2)\gamma^{\alpha}\gamma_5 u_{\chi}(k_1)
=-\frac{k_1^\alpha+k_2^\alpha}{2M_{\chi}}
\bar{v}_{\chi}(k_2)\gamma_5 u_{\chi}(k_1)
-\frac{i}{2M_{\chi}}
\bar{v}_{\chi}(k_2)\sigma^{\alpha\beta}(k_{1\,\beta}-k_{2\,\beta})\gamma_5 u_{\chi}(k_1)
\ee
The vector $(k_1+k_2)^\alpha=(p_1+p_2)^\alpha$ in the  first term saturates the current $\bar u_{f}\gamma_\alpha P_L v_f$ in Eq.~(\ref{currentX})
and gives rise to terms  proportional to the fermion mass, which are zero in our computation.
The second term gives again an $\mathcal{O}(v)$ contribution.
We thus recovered the well-known fact that  for Majorana fermions the scattering amplitude is proportional to the first power of the relative velocity of the incoming particles.
Notice that for Dirac DM  Eq.~(\ref{currentX}) would not contain the
 $D_{21}$ terms, as only the $t$-channel contributes to the amplitude, and the
vector current thus gives rise to an unsuppressed $s$-wave term in the cross section 
(see App.~\ref{app:dirac} for details).


Another interesting limit to analyze   is the large scalar mass regime
 $r\gg 1$ for which
 \be
 D_{ij}\sim \frac{1}{r M_\chi^2}\left[1+{\cal O}\left(\frac{1}{r^2}\right)\right] \quad{\rm and}\quad
 D_{i1}-D_{i2}\sim {\cal O}\left({v\over r^2}\right)\frac{1}{ M_\chi^2}\, .
 \label{di1di2largeMS}
 \ee
In this case, the amplitude for  DM Majorana annihilation into (massless) fermions at leading order in 
$v$ and $1/r$
is given by Eq.~(\ref{currentX}),
where the first term in square brackets is ${\cal O}(v/r^2)/M_\chi^2 $, which is subdominant with respect to the second one, of order $[\mathcal{O}( v/r)/ M_\chi^2] [\bar{v}_{\chi}\; \sigma^{\alpha 3} \gamma_5 u_{\chi}]$; thus, the tree-level cross section will approximately be given by
\be
v\sigma(\chi\chi\to f\bar f)\sim {1\over M_\chi^2}{v^2\over r^2}\,.
\label{estimate2body}
\ee

\section{Three-body DM Annihilation}
\label{sec:3body}

\noindent
Let us now turn to  analyze the case of interest, namely
the emission of EW gauge bosons in DM annihilations.
First, we are going to manipulate the matrix element and discuss
its velocity dependence. Then, we deal with the kinematical constraints
of the 3-body phase space and arrive at the results for the cross section.
Finally,  we re-interpret our findings in the language of
effective field theory  and make some remarks about its use.

\subsection{Matrix element and velocity dependence}
\label{sub:ME}

\begin{figure}[t]
\begin{center}
 \includegraphics[width=8 cm]{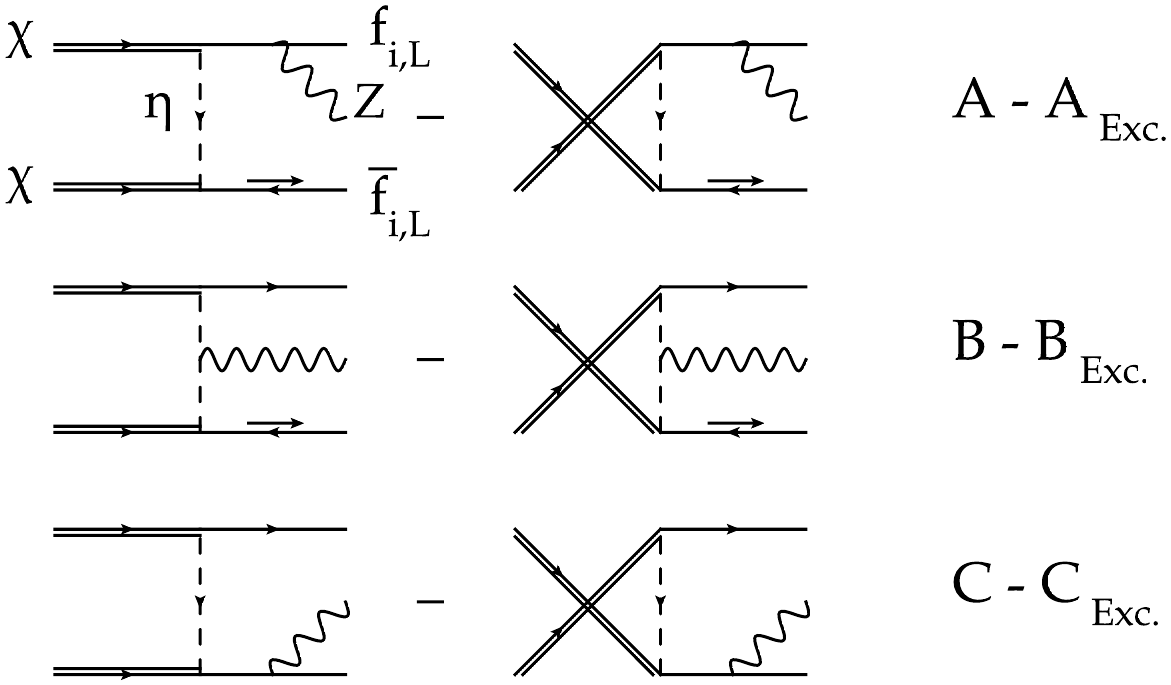}\\
 \caption{\emph{Feynman diagrams for the 3-body process in Eq.~(\ref{eq:3Body}).}}
   \label{fig:3Body}
 \end{center}
\end{figure}

Let us discuss for definiteness the 3-body process with the emission of a $Z$ boson
\begin{equation}\label{eq:3Body}
\chi(k_1)\chi(k_2)\to \bar{f}_L(p_2)f_L(p_1)Z(k)\, .
\end{equation}
The corresponding Feynman diagrams are depicted in Figure \ref{fig:3Body}.
Of course, we shall include also the emission of $W$ gauge bosons in the final results.
The amplitude can be written as
\begin{equation}\label{eq:3BodyAmplitude}
i\mathcal{M}\cdot\epsilon^{*}=\frac{ig|y_L|^2(1-2s_W^2)}{4c_W}\left[
\left(\mathcal{M}_{A}-\mathcal{M}_A^{\rm exc}\right)+
\left(\mathcal{M}_{B}-\mathcal{M}_B^{\rm exc}\right)
+\left(\mathcal{M}_{C}-\mathcal{M}_C^{\rm exc}\right)
\right]\, ,
\end{equation}
where we have denoted $s_W\equiv \sin\theta_W$ and $c_W\equiv \cos\theta_W$, being $\theta_W$ the Weinberg angle.
Following e.g.~Ref.~\cite{bergstrom2}, we shall call ``FSR'' (final state radiation) the processes where a gauge boson is radiated
from an external leg, while we refer to the emission from internal virtual particles as ``VIB'' (virtual internal bremsstrahlung).
Thus, the $A$ and $C$ terms are of FSR type, while the $B$ terms are VIB.

In order to show how the Fierz transformation works, let us analyze in more detail the amplitude $\mathcal{M}_A$,
for massless outgoing fermions
\begin{equation}\label{eq:AmplitudeA}
\mathcal{M}_A=\frac{2[\bar{u}_{f}(p_1)\slashed{\epsilon}^*(k)P_L(\slashed{p}_1+\slashed{k})u_\chi(k_1)]
\left[\bar{v}_\chi (k_2)P_Lv_{f}(p_2)\right]}{(2p_1\cdot k+m_Z^2)(M_{\chi}^2(1-r)-2p_2\cdot k_2)}\, ,
\end{equation}
where each fermionic current is composed by a Dirac and a Majorana spinor, for $f$ and $\chi$ respectively. Applying  the Fierz transformation
\begin{equation}\label{eq:Fierz2}
(P_R)_{ij}(P_L)_{kl}=\frac{1}{2}(P_R\gamma^{\mu})_{il}(P_L\gamma_{\mu})_{kj},
\end{equation}
we can perform the  rearrangement
\bea\label{eq:applicoFierz}
\bar{v}_\chi (k_2)_i(P_L)_{ij}v_{f}(p_2)_j[\bar{u}_{f}(p_1)\gamma^{\rho}(\slashed{p}_1+\slashed{k})]_{k}(P_R)_{kl}u_\chi (k_1)_l\nn\\
=\frac{1}{2}
[\bar{v}_\chi (k_2)P_L\gamma^{\mu}u_\chi (k_1)][\bar{u}_{f}(p_1)\gamma^{\rho}(\slashed{p}_1+\slashed{k})P_R\gamma_{\mu}
v_{f}(p_2)]\, .
\eea
so that Eq.~(\ref{eq:AmplitudeA}) becomes
\begin{equation}\label{eq:AmplitudeAFierzed}
\mathcal
{M}_A=\frac{[
\bar{u}_{f}(p_1)\slashed{\epsilon}^*(k
)(\slashed{p}_1+\slashed{k})P_R\gamma_{\mu}
v_{f}(p_2)]\;[\bar{v}_\chi (k_2)P_L\gamma^{\mu}u_\chi(k_1)]}{(2p_1\cdot k+m_Z^2)(M_{\chi}^2(1-r)-2p_2\cdot k_2)}\, .
\end{equation}
With the same technique, the terms in the total amplitude (\ref{eq:3BodyAmplitude}) read
\begin{eqnarray}
\mathcal{M}_A-\mathcal{M}_A^{\rm exc}&=&
\frac{\bar{u}_{f}\slashed{\epsilon}^*(k)(\slashed{p}_1+\slashed{k})P_R\gamma^{\mu}v_{f}}{2p_1\cdot k+m_Z^2}\cdot \left( \frac{D_{22}- D_{21}}{2}\bar{v}_\chi \gamma_{\mu}u_\chi
+\frac{D_{22}+ D_{21}}{2}\bar{v}_\chi \gamma_{\mu}\;\gamma_5\;u_\chi
\right)\, ,\nonumber\\
\label{eq:A}&&\\
\mathcal{M}_B-\mathcal{M}_B^{\rm exc}&=&
(-1)[\bar{u}_{f}P_R\gamma^{\mu}v_{f}]\left[
(k_1-k_2-p_1+p_2)\cdot\epsilon^*(k)\;\bar{v}_\chi P_L\gamma_{\mu}u_\chi \;D_{1 1}\; D_{2 2}
\right. \nonumber\\
&-&\left.(k_2-k_1-p_1+p_2)\cdot\epsilon^*(k)\;\bar{v}_\chi \gamma_{\mu}P_L u_\chi  \; D_{21}\; D_{12}\;
\right]\,, \label{eq:B}\\
\mathcal{M}_C-\mathcal{M}_C^{\rm exc}&=&-\frac{\bar{u}_{f}P_R\gamma^{\mu}(\slashed{p}_2+\slashed{k})\slashed{\epsilon}^*(k)v_{f}}{2p_2\cdot k+m_Z^2}
\cdot \left( \frac{D_{11}- D_{1\,2}}{2}\bar{v}_\chi \gamma_{\mu}u_\chi
+\frac{D_{11}+ D_{1\,2}}{2}\bar{v}_\chi \gamma_{\mu}\gamma_5 u_\chi
\right)\, .\nonumber\\
\label{eq:C}
&&
\end{eqnarray}
Let us now discuss the  limit $v\ll 1 $ and $r \gg 1 $, in analogy with  the
 previous section for the 2-body process.
 The coefficient of the DM {\it  vector} current in the full amplitudes of the kind  $A$, $B$  and $C$ is always
 ${\cal O}(v)$  as it happens for the 2-body case, due to Eq.~(\ref{di1di2}); in particular,
 in the large $r$ limit it  is proportional to
$ {\cal O}(v/r^2)/ M_{\chi}^2$ (as in Eq.~(\ref{di1di2largeMS})).
Instead, for the {\it axial-vector} current, the crucial point is that
 the mass cancellation of order ${\cal O}(m_f)$ does not  occur anymore.
 Indeed,  from the Gordon identities  
 we recover Eq.~(\ref{GordonId}),
where the second term turns out to be proportional to $v$, while
 in the first term the 4-vector saturating the fermionic currents is now
 $(p_1+p_2+k)_{\mu}$, which does not  trigger anymore the chiral identity, and leaves
 non-zero  terms even for vanishing $m_f$. Thus, the terms of the amplitudes containing $\bar v_{\chi}\gamma_5u_{\chi}$ read
 \be
 \label{ac}
\left.( \mathcal{M}_A-\mathcal{M}_A^{\rm exc}+\mathcal{M}_C-\mathcal{M}_C^{\rm exc}
)\right|_{\bar v_{\chi}\gamma_5 u_{\chi}}
=
[\bar{u}_f\slashed{\epsilon^*}P_L v_f]\frac{[\bar{v}_{\chi}\gamma_5u_{\chi}]}{2M_\chi}\underbrace{\frac{(D_{22}+D_{21})-(D_{11}+D_{12})}{2}}_{{\cal O}\left({1\over r^2}\right)
\frac{1}{M_\chi^2}}\\
 \ee
\begin{eqnarray}
 \nonumber\label{bb}
\left.( \mathcal{M}_B-\mathcal{M}_B^{\rm exc} )\right|_{\bar v_{\chi}\gamma_5u_{\chi}}&=&
-[\bar{u}_f\slashed{k}P_L v_f]\frac{[\bar{v}_{\chi}\gamma_5u_{\chi}]}{2M_\chi}\;
\bigg[
(p_2-p_1)\cdot\epsilon^*(k)  \;\underbrace{\frac{(D_{11}\;D_{22}+D_{12}\;D_{21})}{2}}_{{\cal O}\left({1\over r^2}\right)\frac{1}{M_\chi^4}}
+
\\ &&
\underbrace{(k_1-k_2)\cdot\epsilon^*(k)}_{vM_{\chi}\epsilon^*_z} \;\underbrace{\frac{(D_{11}\;D_{22}-D_{12}\;D_{21})}{2}}_{{\cal O}\left({v\over r^3}\right)\frac{1}{M_\chi^4}}
 \bigg]\,,
\end{eqnarray}
where we have highlighted the behavior of each term with $v$ and $1/r$.
Notice that now there appear terms without $v$ dependence. Indeed, in the limit $v=0$ and to leading order in $1/r\ll 1$, the full amplitude is given by
 \be \left.{\cal M}\right |_{v\to 0}=
\frac{\left(\bar{v}_{\chi}\gamma_5u_{\chi}\right)}{2M_\chi^5} \frac{1}{r^2}
\left[(\bar{u}_f\slashed{\epsilon^*}P_L v_f) \;(p_1-p_2)\cdot (k_1+k_2)-
\left(\bar{u}_f\slashed{k}P_L v_f\right)\;\;(p_2-p_1)\cdot \epsilon^*\right],
 \ee
where the first term comes from FSR while the second originates from VIB, and
they are both of order $\mathcal{O}(1/r^2)$.
Schematically, the various contributions to the amplitude can be organized as follows
\be
{\cal M}\sim {1\over M_\chi}{\cal O}(v)\left [ \left.{\cal O}\left({1\over r}\right)\right\vert_{\rm FSR}
+ \left. {\cal O}\left({1\over r^2}\right)\right\vert_{\rm FSR} \right]+
{1\over M_\chi}
\left[  \left.{\cal O}\left({1\over r^2}\right)\right\vert_{\rm VIB} + \left.{\cal O}\left({1\over r^2}\right)\right\vert_{\rm FSR} \right]
\,.
\label{Mestimate}
\ee
At this point we can learn an important lesson (see also Ref.~\cite{bergstrom1}):
the opening of the $s$-wave originates from diagrams of both FSR and VIB type, at  $\mathcal{O}(1/r^2)$ in the amplitude; limiting the expansion up to $\mathcal{O}(1/r)$ in the amplitude would cause
the process to stay in the $p$-wave.

An order-of-magnitude estimate  for the 3-body cross section,
showing the leading dependence on the expansion parameters,
can be obtained straightforwardly
\be
v\sigma(\chi\chi\to f\bar f Z)\sim \frac{\alpha_W}{M_{\chi}^2}\left[
 {\cal O}\left({v^2\over r^2}\right)+
{\cal O}\left({v^2\over r^3}\right)+
{\cal O}\left({1\over r^4}\right)
\right]\, ,
\label{estimate3body}
\ee
where  the weak coupling  $\alpha_W=g^2/(4\pi)$
for the gauge boson emission has been restored.
The estimates in Eqs.~(\ref{estimate2body}) and (\ref{estimate3body}) allow to gather an understanding in simple terms  of the
situation we are studying.
While the 2-body annihilation cross section behaves like $v^2/r^2$, the 3-body  FSR and VIB diagrams give rise
to both $s$-wave and $p$-wave terms. 
The $p$-wave from 3-body processes cannot compete with the 2-body cross section because of the
extra $\alpha_W$ factor; however the $s$-wave from $2\to 3$ annihilation, free from the $v^2$ suppression, can overcome the $2 \to 2$
cross section
if $r$ is not too large. In the next subsection we shall give a more precise estimate based on the analytical results.

Because of the importance of this point, and being the distinction between FSR and VIB not able to disentangle clearly the $s$-wave contribution from the $p$-wave one, let us introduce now a specific notation. Having in mind an expansion in powers
of $1/r$ in the amplitude -- as sketched in Eq.~(\ref{Mestimate}) -- we shall call ``leading order'' (LO) the lowest order
term $\mathcal{O}(1/r)$ in this expansion, which originates from lowest order FSR-type diagrams.
This is the order at which Refs.~\cite{paper1, w} work. 
As shown above, in the LO approximation the annihilation cross section proceeds through $p$-wave.  Only 
higher order terms are able to remove the helicity suppression.
We shall further elaborate on this expansion as an operator expansion in Sect.~\ref{sec:effective}.

As a check of the results of this subsection, one can use the Ward Identities for EW SM gauge bosons
$k_{\mu}{\cal M}_L^{\mu}\sim 0$, for  $m_f\sim 0$,
where ${\cal M}_L^{\mu}$ is the amplitude computed for the longitudinal mode of the $Z$.
By direct  calculation one  obtains
\bea
k_{\mu}({\cal M}^{\mu}_{ A}-{\cal M}^{\mu\;{\rm exc}}_{ A})&=& (\bar{u}_f\gamma_{\alpha}P_Lv_f)
\left[\frac{D_{22}- D_{21}}{2}
(\bar{v}_\chi \gamma^{\alpha}  u_\chi ) +
 \frac{D_{22}+D_{21}}{2}
( \bar{v}_\chi \gamma^{\alpha}\gamma_5 u_\chi)
\right]\, ,\\
k_{\mu}({\cal M}^{\mu}_{ C}-{\cal M}^{\mu\;{\rm exc}}_{ C})&=& -(\bar{u}_f\gamma_{\alpha}P_L v_f)
\left[\frac{D_{11}-D_{12}}{2}
(\bar{v}_\chi \gamma^{\alpha}  u_\chi ) +
 \frac{D_{11}+ D_{12}}{2}
( \bar{v}_\chi \gamma^{\alpha}\gamma_5 u_\chi)
\right]  \\
k_{\mu}({\cal M}^{\mu}_{ B}-{\cal M}^{\mu\; {\rm exc}}_{ B})&=& -(\bar{u}_f\gamma_{\alpha}P_L v_f)
\Big[
[D_{22}-D_{11}-(D_{21}-D_{12})]
(\bar{v}_\chi \gamma^{\alpha}  u_\chi )+ \nonumber \\
&+&
\left[D_{22}-D_{11}+(D_{21}-D_{12})\right]
(\bar{v}_\chi \gamma^{\alpha} \gamma_5 u_\chi  )
\Big] \, ,
\eea
whose vanishing sum  confirms the Ward Identity
\footnote{In the ${ B}$ diagrams, we used the trick
$(k_1-k_2+p_2-p_1)\cdot k=D_{11}^{-1}-D_{22}^{-1}$ and $(k_2-k_1+p_2-p_1)\cdot k=D_{12}^{-1}-D_{21}^{-1}$.}.

It is interesting to see the level of cancellation in the large $M_S$ limit using the properties of Eq.~(\ref{dd}).
We see that, up  to order ${\cal O}(1/r)$,  ${\cal M}_A$ and  ${\cal M}_C$ cancel each other so that, at this order,
we can say that their sum is gauge invariant;  if we want to keep corrections of order ${\cal O}(1/r^2)$ or higher, the full sum of ${ A,C}$ and ${ B}$  diagrams have to be considered in order to have a consistent evaluation
\be\label{WI}
k_{\mu}(\mathcal{M}^\mu_A-\mathcal{M}_A^{\mu \,\rm exc}+\mathcal{M}^\mu_C-\mathcal{M}_C^{\mu \,\rm exc})
={\cal O}\left({1\over r^2}\right)+ \cdots=-k_{\mu}(\mathcal{M}^\mu_B-\mathcal{M}_B^{\mu \,\rm exc}).
\ee
If we do not sum up the full corrections, after summing over the polarizations of the
outgoing massive vector,
we would  end up with unphysical (non-decoupling) $(M_{S}/m_Z)^2$ and $(M_{\chi}/m_Z)^2$ corrections \cite{Ciafaloni:2010qr}.

\subsection{Results for the cross section}
\label{subsec:xsec}

We now turn to  evaluate the full 3-body cross section for the process in Eq.~(\ref{eq:3Body}), including VIB diagrams.
We follow a rather pedagogical approach, starting from the formula for the cross
section
\begin{equation}\label{eq:sezione}
d\sigma=\frac{|\mathcal{M}|^2}{4\mathcal{I}}(2\pi)^4\delta^{(4)}(k_1+k_2-p_1-p_2-k)\frac{d\textbf{p}_1}{(2\pi)^32p_1^0}
\frac{d\textbf{p}_2}{(2\pi)^32p_2^0}\frac{d\textbf{k}}{(2\pi)^32k^0},
\end{equation}
being $\mathcal{I}=[(k_1\cdot k_2)^2-M_{\chi}^4]^{1/2}$ the initial flux.
The squared amplitude $|\mathcal{M}|^2$ is obtained from Eqs.~(\ref{eq:3BodyAmplitude}) and (\ref{eq:A})-(\ref{eq:C}) by summing over
  the physical gauge boson polarizations
\begin{equation}\label{pol}
\sum_{i=1,2,3}\epsilon^i_\mu(k)\epsilon^{i*}_\nu(k)=-g_{\mu\nu}+\frac{k_\mu k_\nu}{m_Z^2}\,.
\end{equation}
Integrating Eq.~(\ref{eq:sezione}) over the three angles that define the position of the plane described through the momentum conservation $\textbf{p}_1+\textbf{p}_2+\textbf{k}=0$ we obtain
\begin{equation}\label{eq:3BodyCrossSectionFormula}
vd\sigma=\frac{|\mathcal{M}|^2}{1024\pi^4}dx_1dx_2\, ,
\end{equation}
where $x_1$ and $x_2$  parametrize the final energies. In particular, letting $s_1\equiv (k_1+k_2)^2$,
we have
\begin{eqnarray}
 k^0 &=& (1-x_2)\sqrt{s_1}/2\, , \label{eq:k}\\
 p_1^0 &=& x_1\sqrt{s_1}/2\, ,\label{eq:p1} \\
 p_2^0 &=& (1-x_1+x_2)\sqrt{s_1}/2\, ,\label{eq:p2}
\end{eqnarray}
and we find the following constraints on the phase space
\begin{eqnarray}
 x_{-}\leq &x_1& \leq x_+ \quad {\rm with}\quad
x_{\pm}=
{1+x_2\over 2}\pm\sqrt{{(1-x_2)^2\over 4}-\frac{m_Z^2}{s_1}}\quad
\\
-\frac{m_Z^2}{s_1}\leq & x_2&\leq 1-2\;\frac{m_Z}{\sqrt{s_1}}\,.
\label{eq:phasespace}
\end{eqnarray}
The integrations of the squared amplitude over the phase space cannot be carried out exactly, but
two  limiting situations are of interest: an  expansion in powers of $1/r\ll 1$,
and the case with $v\to 0$ with $r$ generic. The results in the former limit are shown below, while
the latter case is reported in Appendix \ref{app:vzero}.

Let us parametrize the cross section as
\begin{equation}\label{eq:TotalCrossSection3Body}
v\sigma=\frac{\alpha_W|y_L|^4(1-2s_W^2)^2}{64\pi^2 c_W^2 M_\chi^2}\left(\rho_s+\rho_p\right)\, .
\end{equation}
The partially-inclusive cross section, expanded in the large $M_S$ limit $(r\gg 1)$,
is obtained  by integrating Eq.~(\ref{eq:3BodyCrossSectionFormula}) over $x_1$; 
neglecting terms vanishing in the $m_Z\to 0$ limit we find
\bea
\label{riss}
{d\rho_s\over d x_2}&=&\frac{4}{3r^4} x_2 (1-x_2)^3\;+\mathcal{O}(r^{-5})\, ,\\
{d\rho_p\over d x_2}&=&\frac{v^2}{3r^2}\,\left[
  \frac{1+x_2^2}{1-x_2}
\left(1-{4\over r}+{11\over r^2}\right)\log\frac{\bar x_+}{\bar x_{-}}
\right.\nn\\
\label{risp}
&&\left.
+  \left(1-x_2\right)\left[
2- {2\left(x_2+5\right) \over r}+
{\left(-3x_2^3+14x_2^2+17x_2+34\right)\over r^2}\right]\right]
\;+\mathcal{O}(r^{-5})\, .
\eea
with $\bar x_{\pm}=1-x_2\pm \sqrt{(1-x_2)^2-\frac{m_Z^2}{M_{\chi}^2}}$.

The fully-inclusive $s$-wave and $p$-wave contributions are
obtained by further integrating over $x_2$ as prescribed in Eq.~(\ref{eq:phasespace}); in the large $r$ limit, we get
\bea
\label{eq:SwaveFinale}
\rho_s&=&\frac{1}{15 r^4 }\;+\mathcal{O}(r^{-5})\, ,\\
\rho_p&=&\frac{v^2}{180 r^2}\left[
60\left(1-{4\over r}+{11\over r^2}\right)\ln\frac{2M_{\chi}}{m_Z}\left(2\ln\frac{2M_{\chi}}{m_Z}-3\right)
\right.\nonumber\\
&&\left.
+10(15-\pi^2)
+{40(\pi^2-13)\over r}
+{(1059-110\pi^2)\over r^2}\right]
\;+\mathcal{O}(r^{-5})\, .
\label{eq:PwaveFinale}
\eea
This result shows explicitly the peculiar structure in powers of $1/r$ that we have estimated in Eq.~(\ref{estimate3body}) using general arguments.
As an order-of-magnitude estimate, for velocities $v\sim 10^{-3}$, one expects the 3-body process (in $s$-wave) to dominate over the 2-body one (in $p$-wave)  for values of $r\lesssim \mathcal{O}(10)$.

Our results 
are in disagreement with the ones of
  Ref. \cite{w}, where the helicity suppression was removed just by including LO FSR, and 
  where in addition EW corrections proportional to
$M_\chi^2/m_Z^2$ were found.
Indeed, such terms are present in the physical
  polarizations sum in Eq.~(\ref{pol}). 
 However, because of gauge invariance in
  the form of the Ward Identities written in Eq.~(\ref{WI}), they disappear from the final
  result as they should.
On the contrary, we find a perfect agreement with the result of Ref.~\cite{bergstrom1} where the photino annihilation process 
$\widetilde{\gamma}\widetilde{\gamma}\to e^+e^-\gamma$ in the $v\to 0$ limit is analyzed.

\subsection{Effective Field Theory Approach}
\label{sec:effective}

Let us now see how the previous results can be interpreted  in the effective field theory language.
If the mass of the intermediate scalar particle is much larger than the energy scale of the
non-relativistic annihilation  ($E\simeq M_\chi$), then it is possible to
 integrate the scalar out and perform an operator expansion in the small parameter $1/r=M_\chi^2/M_S^2$.
 The effective Lagrangian will be given by an infinite  series
\be
{\cal L}_{\rm{eff}}={\cal L}_{\rm{SM}}+{\cal L}_{\chi}+{1\over r}{{\cal O}_{6}\over M_\chi^2}+{1\over r^2}
{{\cal O}_{8}\over M_\chi^4}+...\,,
\ee
where ${\cal O}_{n}$ are dimension-$n$ operators.
For the theory we considered in Eq.~(\ref{eq:Lagrangian}), the single  dimension-six operator is
\begin{equation}\label{eff}
{\cal O}_{6}=
\frac{1}{2}|y_L|^2
\left[\bar{\chi}\gamma_\mu \gamma_5\chi \right]
\left[\bar L\gamma^\mu P_L L\right]\,,
\end{equation}
which generates the tree level contributions
of Eq.~(\ref{currentX}).
The corresponding cross section for the $\chi\chi\to f\bar f$ process is
\begin{equation}
\left.
v\sigma(\chi\chi\to f\bar f)\right\vert_{{\cal O}_6}=\frac{|y_L|^4 }{48\pi M_\chi^2 }{v^2\over r^2 }\,,
\end{equation}
where the usual $v^2$-suppression appears. Notice that this result can be recovered from
Eq.~(\ref{bpwave}) in the  limit $r\gg 1$ 
and corresponds to the estimate in  Eq.~(\ref{estimate2body}).

As soon as we add  the $Z$  emission (for simplicity we consider only the presence of a
single gauge boson $Z$),  at the LO ${\cal O}(1/r)$ in the amplitude
only the external leg can radiate a gauge boson, making diagrams of the FSR type.
We have performed a complete calculation using these effective amplitudes.
For the total 3-body cross section we find, neglecting terms vanishing in the limit $m_Z\to 0$
\begin{equation}\label{eq:3BodyBremmTotal}
\left.v\sigma(\chi\chi\to f\bar f Z)\right\vert_{{\cal O}_6}=
\frac{\alpha_W|y_L|^4(1-2s_W^2)^2}{1152\pi^2 c_W^2 M_\chi^2}{v^2\over r^2}\left[
15-\pi^2+6\ln\frac{2M_{\chi}}{m_Z}\left(2\ln\frac{2M_{\chi}}{m_Z}-3\right)
\right]\; .
\end{equation}
This expression still bears the $v^2$ dependence, so that one recovers  the result that the EW radiation from 
the external legs cannot remove the $p$-wave suppression at LO.
Notice that Eq. (\ref{eq:3BodyBremmTotal}) exhibits the usual single and double
logarithmic behavior of infrared origin, and that it correctly reproduces 
the limit for $r\gg 1$
of Eqs.~(\ref{eq:TotalCrossSection3Body}) and (\ref{eq:PwaveFinale}).

From these results it is clear that limiting the analysis to the dimension-six
operator in the effective theory, which corresponds to work in the LO approximation, 
misses the right result since one could
incorrectly conclude that the total cross section itself is $p$-wave suppressed.
As shown in the previous section instead, in order to obtain the
correct result, namely that the cross section receives important $s$-wave contributions,
one needs to consider the diagrams (of VIB and FSR type) arising
at the next order.
Therefore the effect of removing the suppression is
encoded by operators of dimension higher than six, for example
those in $\mathcal{O}_{8}$.

\section{Energy spectra of final stable particles at the interaction point}
\label{sec:spectra}

The analytical results obtained above have a phenomenological impact for
DM indirect searches. Indeed, the energy spectra of  stable particles  
produced by DM annihilation, with the inclusion of EW bremsstrahlung,
can be very different from those commonly obtained by working at the LO.
In this section we show our results for the energy spectra at the interaction point,
focusing in particular on positrons, antiprotons, photons and neutrinos.
The propagation of these fluxes of stable particles  through the galactic halo
will be discussed in Section \ref{sec:propagation}.
Our analysis is based on the combination of the analytical description of the primary annihilation channels with the numerical techniques  for subsequent hadronization and decay.
Let us now describe our procedure in more detail.

As already discussed, we work in the approximation of massless external fermions, under which the calculations of the previous sections have been performed.
So we consider  only the case  where $L=( \nu_{e\,L}, e_L)$, for which this is an excellent approximation.
In general, channels with external fermions of mass $m_f$ would receive other, different contributions
proportional to $(m_f/M_\chi)^2$, in addition to the $s$-wave contribution from VIB and FSR, as discussed
at length above.
The  primary annihilation channels for $\chi\chi\to I$, including EW bremsstrahlung, are
\begin{equation}
I= \{ e^+_L e^-_L,\, {\nu}_{e\,L}\bar {\nu}_{e\,L},\,
e^+_L e^-_L\gamma, \, e^+_L {\nu}_{e\,L} W^-,\, e^-_L\bar{\nu}_{e\,L}\, W^+, \,
e^+_L e^-_L Z, \, {\nu}_{e\,L}\bar{\nu}_{e\,L} Z
\}
 \,.
  \label{eq:channels}
\end{equation}
The different 3-body channels are simply related by  different gauge couplings
\begin{eqnarray}
  \sigma(\chi\chi\to {\nu}_{e\,L}\bar{\nu}_{e\,L}Z) &=& \frac{1}{(1-2s_W^2)^2}\sigma(\chi\chi\to e^+_Le^-_LZ), \\
  \sigma(\chi\chi\to e^-_L\bar {\nu}_{e\,L} W^+)= \sigma(\chi\chi\to e^+_L {\nu}_{e\,L} W^-) &=&  \frac{2c_W^2}{(1-2s_W^2)^2}\left.\sigma(\chi\chi\to e^+_Le^-_LZ)\right|_{m_Z\to m_W},\\
   \sigma(\chi\chi\to e^+_Le^-_L\gamma) &=& \frac{4c_W^2s_W^2}{(1-2s_W^2)^2}\left.\sigma(\chi\chi\to e^+_Le^-_LZ)\right|_{m_Z\to 0}.
\end{eqnarray}
We have written our own Monte Carlo code to generate parton-level
events for DM annihilations into 2- and 3-body final states,
in the frame where the total spatial momentum is zero.
While for the $2\to 2$ processes the final state  consists of two back-to-back particles,
 the 3-body final states require particular care because the probability distribution of the
 momenta of the outgoing particles
is dictated by  the double-differential probability distributions
\begin{equation}\label{eq:doublediff}
\frac{1}{\sigma(\chi\chi\to 3{\rm -body})}\frac{d\sigma(\chi\chi\to 3{\rm-body})}{dx_1dx_2},
\end{equation}
where $x_1$ and $x_2$ are related to the energy fractions of the outgoing particles,
as in Eqs. (\ref{eq:k}), (\ref{eq:p1}), (\ref{eq:p2}).

A  large number 
of events ($2\times 10^5$) for each annihilation channel in Eq.~(\ref{eq:channels}) is generated in this way,
and then passed through {\sc Pythia} 8.145 \cite{pythia} for simulating the subsequent showering,
hadronization and decay
\footnote{{\sc Pythia} 8.1  has been preferred over the  predecessor {\sc Pythia} 6.4
because of the inclusion of the photon branchings into fermion-antifermion in the showering process.}.
All unstable particles are requested to decay so that the only
 final particles remaining in the sample are the stable species of the SM.
We have performed several checks at various levels to
assess the reliability of our numerical code.
For instance, we have found excellent agreement with the results of  Ref.~\cite{paper1}.

\begin{figure}[t]
\begin{center}
 \includegraphics[scale=1]{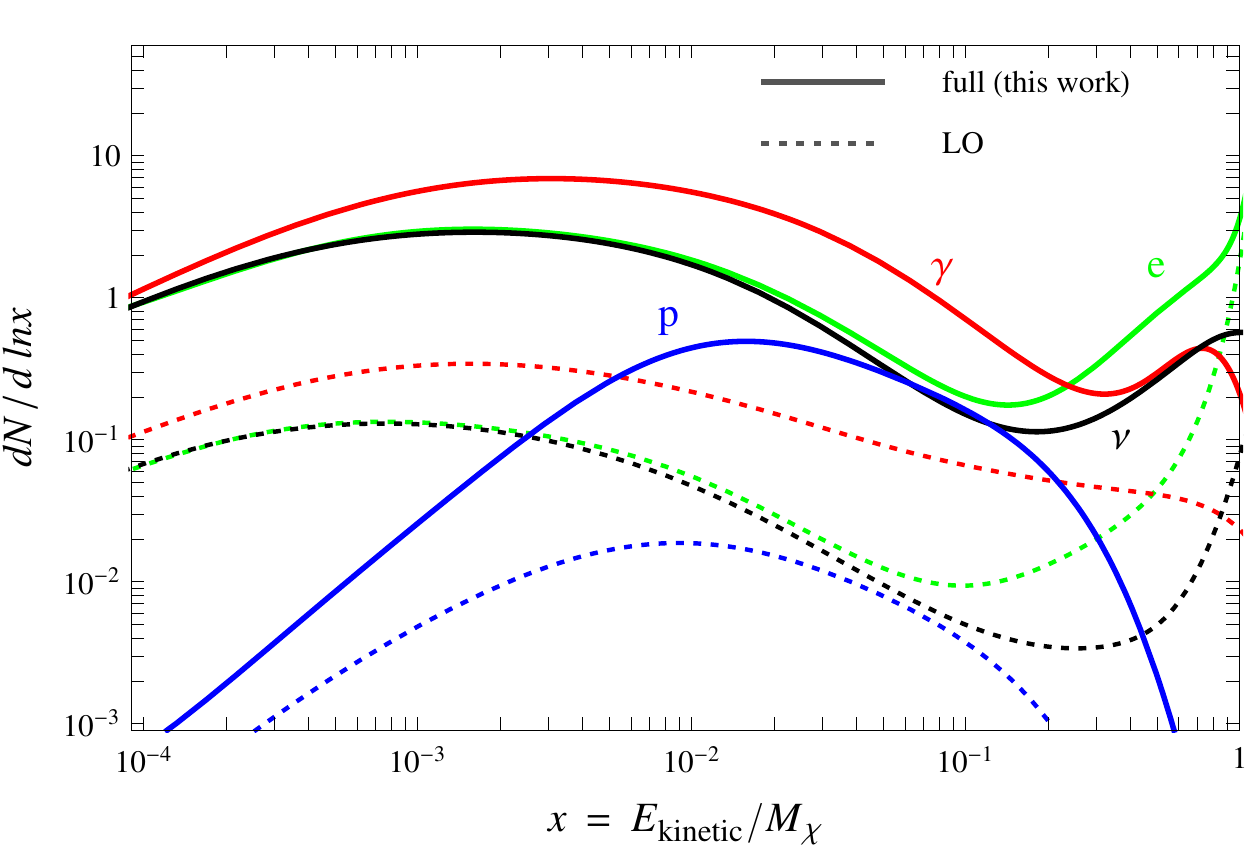}
 \caption{\emph{The spectra ${d{\cal N}_f}/{d \ln x}$, as defined in Eq.~(\ref{eq:norm}), for
  $e^+$ (green), $\gamma$ (red), $\nu=(\nu_e+\nu_\mu+\nu_\tau)/3$ (black)
 and $\bar p$ (blue),  from the annihilation $\chi\chi\to e_L^+ e_L^-, \nu_{e\, L}\bar{\nu}_{e\,L}$
 with the corresponding weak boson emission corrections, for the case  $M_\chi=1 \TeV, M_S=4 \TeV, v=10^{-3}$ \emph{(solid lines)}.
 For comparison, we show  the spectra  \emph{(dashed lines)} in the LO approximation (see text for details).}
 }
 \label{fig:spectra}
 \end{center}
\end{figure}

Fitting the numerical results, it is possible to extract the energy distributions
of  each stable particle $f$
\be
\label{eq:norm}
\frac{d{\cal N}_f}{d \ln x}\equiv {1\over \sigma_0}{d\sigma(\chi\chi\to f+X)\over d\ln x}
 \,,\qquad f=\{ e^+, e^-, \gamma, \nu,\bar\nu, p,\bar p\}\,,
\ee
where $x\equiv E_{\rm kinetic}^{(f)}/M_{\chi}$, $E_{\rm kinetic}^{(f)}$ is the kinetic energy of the particle $f$ (the difference between total and kinetic energies is obviously relevant only for the (anti)protons), and the  $X$ reminds us of the inclusivity in the final state with respect to the particle $f$.
As a normalization, we have chosen the tree-level cross section of the 2-body processes
\footnote{
Another choice for the normalization would be the total cross section 
$\sigma(\chi\chi\to f+X)$, which would provide the quantity in Eq.~(\ref{eq:norm}) with a more transparent physical
interpretation as the energy spectrum of $f$.
However,  $\sigma(\chi\chi\to f+X)$ is not as easily calculable as the 2-body cross section and it would not be 
possible to compare the energy spectra 
with and without the $s$-wave contributions because their total cross sections would be different.
In any case, the specific choice of the normalization becomes  irrelevant when taking ratios of spectra,
which serve to stress the relevance of the effect we are studying.
}
\begin{equation}\label{eq:set}
\sigma_{0}=\sigma_{\rm tree}(\chi\chi\to e^+_Le^-_L)+\sigma_{\rm tree}(\chi\chi\to \nu_{e\,L}\bar{\nu}_{e\,L})\,.
\end{equation} 

The plot  in Figure \ref{fig:spectra} shows  the resulting ${d{\cal N}_f}/{d x}$  for $e^+, \gamma, \nu=(\nu_e+\nu_\mu+\nu_\tau)/3, 
\bar p$ for a specific, but representative, choice of parameters:
$M_\chi=1 \TeV$, $M_S=4 \TeV$ (corresponding to $r=16$) and  $v=10^{-3}$;
for comparison, the situation where only the LO term is taken into account is also shown (recall
that what LO means has been discussed in Sect.~\ref{sub:ME}). 
The  energy spectra  result to be much larger than those obtained in the LO approximation.
This is a consequence of having a sizeable $s$-wave annihilation channel opened at 
the next-to-leading order in the $1/r$ expansion.
To better clarify this point we remind that -- as already discussed in Ref.~\cite{paper1} -- the emission of an EW gauge boson opens the hadronic channel, and has dramatic consequences on the final state: antiprotons as well as a large number of soft photons, positrons and neutrinos from pion decays are produced leading to a huge enhancement in the low-energy tail of the energy spectra of final stable particles. The situation described in Ref.~\cite{paper1} is obtained here in correspondence of the LO approximation where - as discussed in Section \ref{sec:3body} - the $p$-wave term in the 3-body cross section dominates widely over the $s$-wave one, giving corrections factorized with respect to the tree-level annihilation process. 
Going beyond the LO, the opening of a  sizeable $s$-wave contribution results into a harder energy spectrum for the primary gauge boson which is entirely
converted (after decay and hadronization processes)
into low energy stable particles, thus leading to a greater
enhancement in the low-energy tails of their spectra.

\begin{figure}[t]
      \centering
       \includegraphics[scale=0.8]{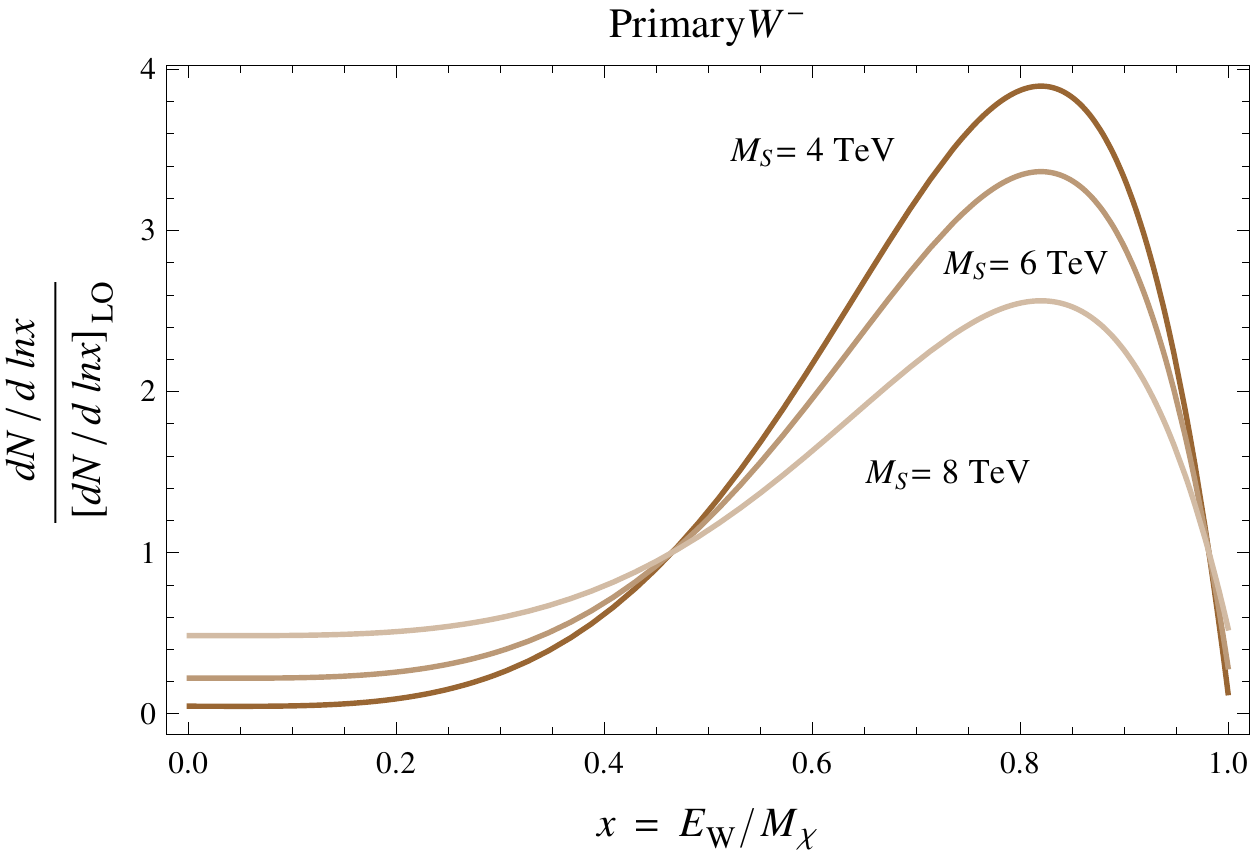}
\caption{\emph{Ratios between the energy spectrum of the $W^{-}$ gauge boson in the 3-body annihilation channel $\chi\chi\to e^+_L\nu_{e\,L}W^-$ for different values of $M_S$ with respect to the same energy spectrum computed in the LO approximation (see text for details).}}
\label{fig:primaryratios}
   \end{figure}

An interesting spectral feature of the gamma rays originates from the inclusion of EW corrections 
(see Figure \ref{fig:spectra}). Indeed, the gamma ray spectrum is the composition of a bump in the hard region due to the contribution of hard photons coming from the $s$-wave in the primary annihilation channel $\chi\chi\to e^+_Le^-_L\gamma$, and a huge tail of soft ones originating from showering processes and from the hadronization of the $W$ and $Z$ gauge bosons included in our analysis.

\begin{figure}[t]
  \begin{minipage}{0.4\textwidth}
   \centering
   \includegraphics[scale=0.6]{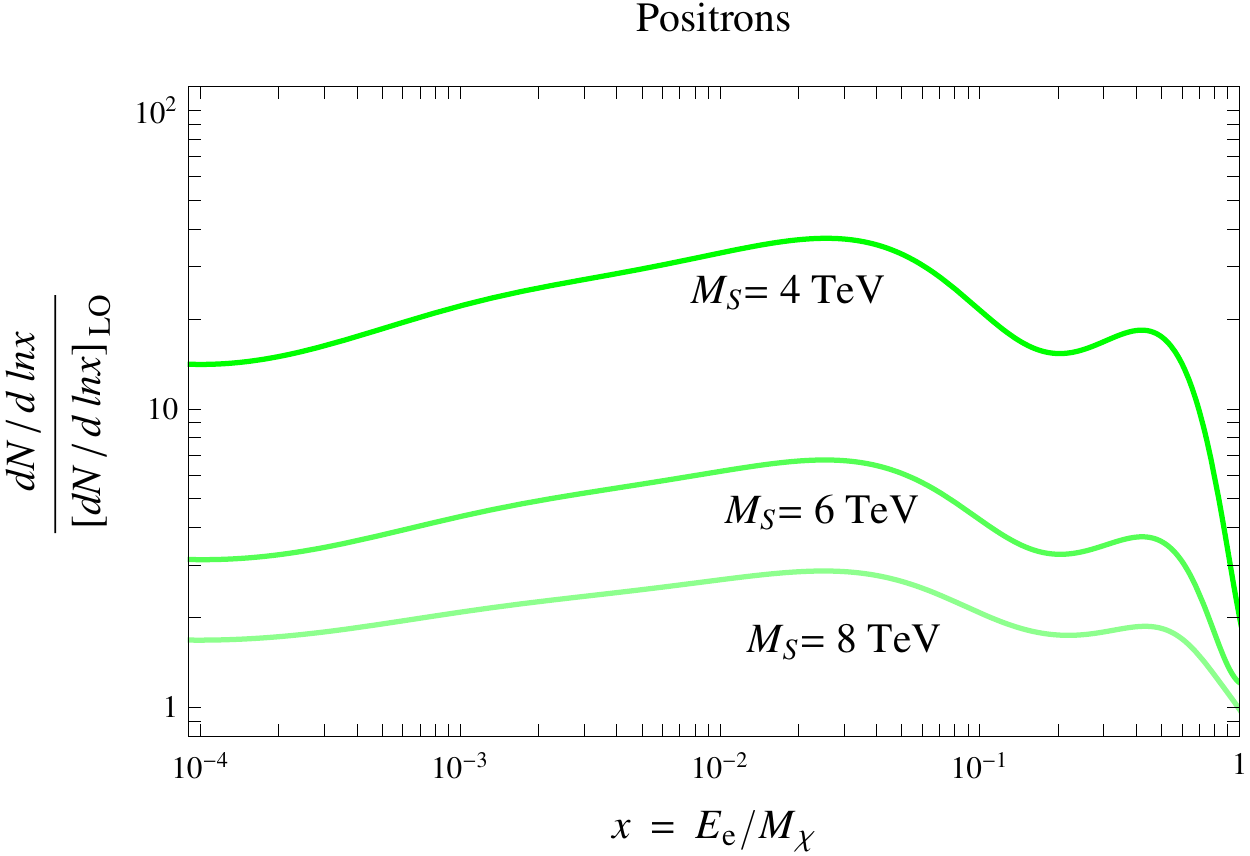}\\
   \vspace{0.5cm}
   \includegraphics[scale=0.6]{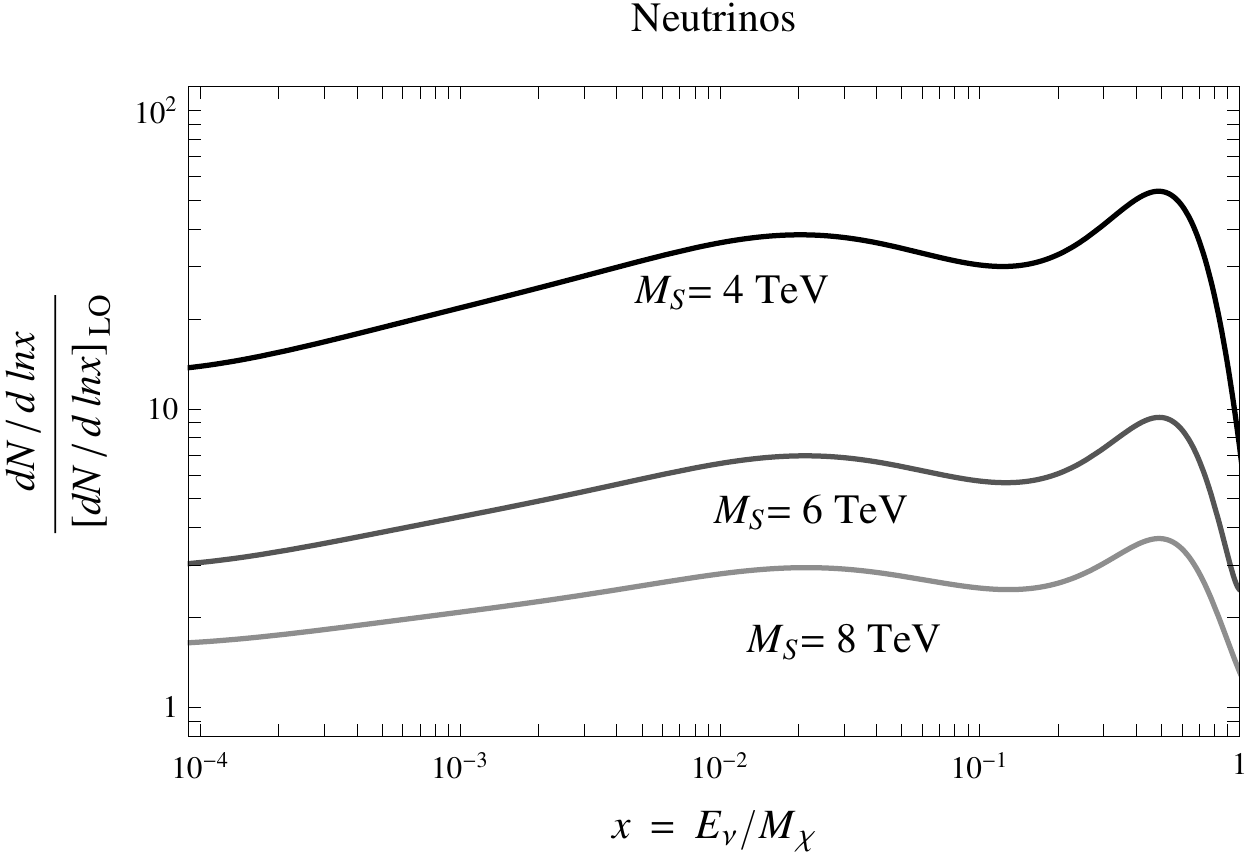}
    \end{minipage}\hspace{1.5 cm}
   \begin{minipage}{0.4\textwidth}
    \centering
    \includegraphics[scale=0.6]{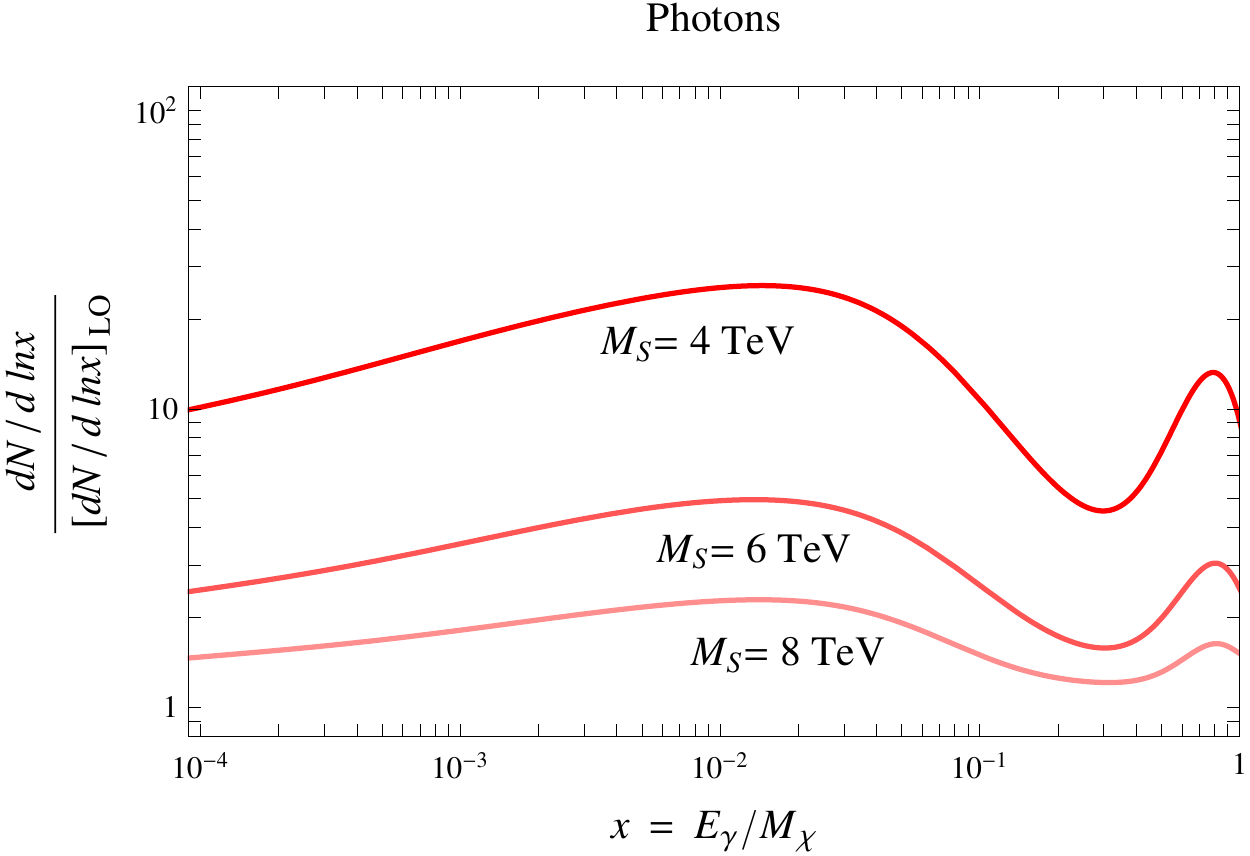}\\
      \vspace{0.5cm}
   \includegraphics[scale=0.6]{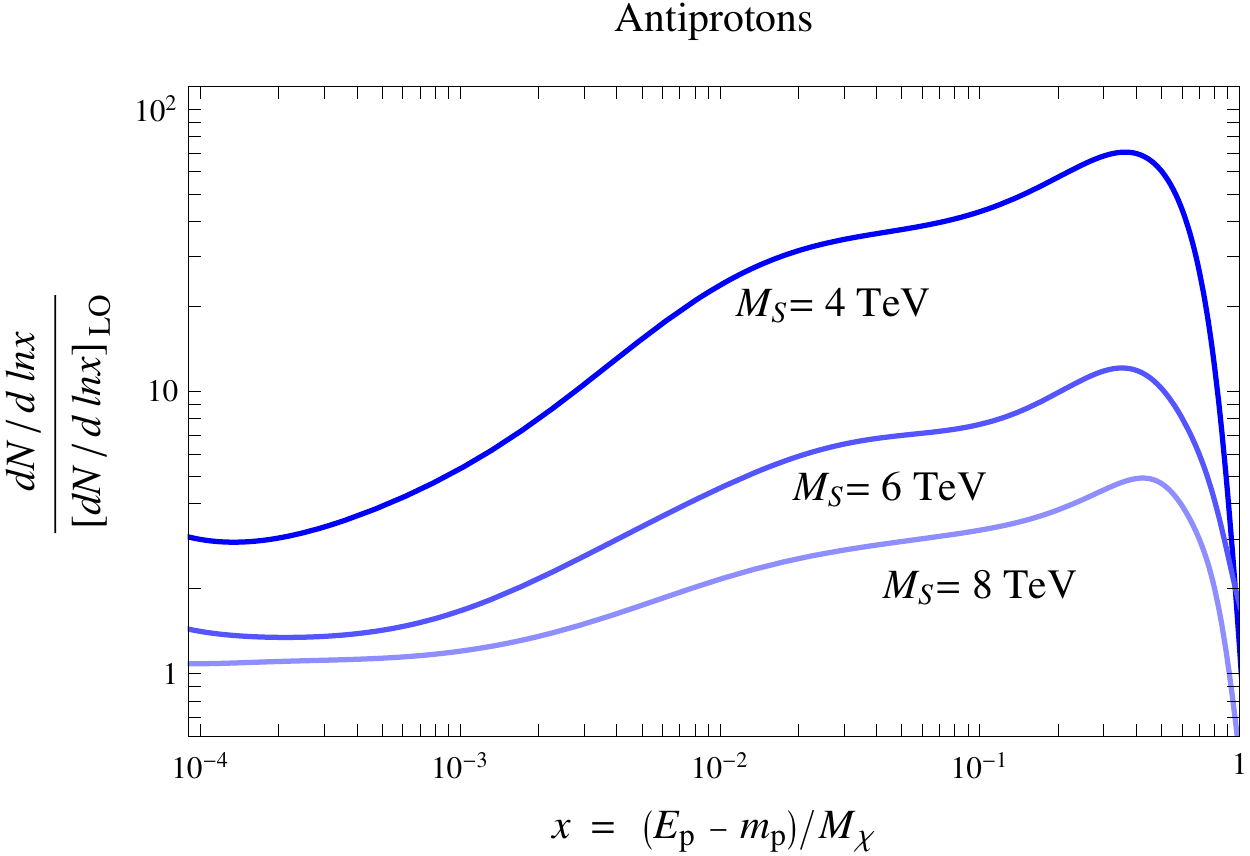}
    \end{minipage}
 \caption{\emph{Ratios between the energy spectra of final stable particles for different values of $M_S$ with respect to those computed in  the LO approximation (see text for details).}}
 \label{fig:ratios}
\end{figure}

The relevance of the effect of removing the suppression  is made more manifest by taking the
 ratios between the energy spectra computed at given $M_S$ with respect to those obtained
 in the LO approximation, as shown
  in Figure \ref{fig:primaryratios} and in Figure \ref{fig:ratios}, for different values of $M_S$.
 In Figure \ref{fig:primaryratios} we plot these ratios before hadronization and decays, considering as example the 
 spectrum of the primary $W^-$ gauge boson in the 3-body annihilation channel
 $\chi\chi\to e^+_L\nu_{e\,L}W^-$. The growth in the hard region as the $s$-wave contribution becomes larger is apparent; this prerogative is present in all the 3-body channels included in our analysis and listed in Eq.~(\ref{eq:channels}).
  In Figure \ref{fig:ratios}, we show the ratios of the energy spectra of the final stable particles
  after  hadronization and decays, with respect to those at LO.
The impact of the EW radiation beyond LO can result into an enhancement of the
energy spectra even by factors $\mathcal{O}(10-100)$.

\section{Fluxes of final stable particles at detection}
\label{sec:propagation}

The previous sections have focused on the calculations of the energy spectra of stable SM particles at the interaction point, normalized for each DM annihilation event. In this section we want to make contact with the phenomenological observables and thus compute the fluxes of electrons, positrons, antiprotons, prompt gamma rays and neutrinos that can be measured at Earth. We first recall briefly the basics of the computation of such fluxes (see e.g.~Ref.~\cite{SalatiCargese} for a lucid and pedagogical review) and then illustrate the results for the energy spectra found in the previous section.

\subsection{Basics of galactic propagation of stable particles}

\noindent {\bf Dark Matter distribution in the galactic halo.} The DM density profile in the galactic halo, $\rho(\vec x)$, is one of the essential ingredients to determine the normalization of the fluxes of cosmic rays that are collected at Earth.
N-body numerical simulations performed in the latest decades have found different answers for $\rho(r)$. While recent simulations seem to individuate the Einasto profile as the best option, the Navarro-Frenk-White (NFW) profile is stil widely used in the literature and the cored Burkert profile (disfavored by simulations) is sometimes advocated as a better fit to astronomical observations. These profiles explicitly read
\begin{center}
\footnotesize{
\begin{tabular}{c|c|c|c}
 & $\rho(r)  $ & $r_{s}$ [kpc] & $\rho_{s}$ [GeV/cm$^{3}$]  \\
\hline
& & \\[-10pt]
NFW~\cite{NFW} & $\displaystyle \rho_{s}\frac{r_{s}}{r}\left(1+\frac{r}{r_{s}}\right)^{-2}$ & 24.42 & 0.184  \\
& & \\[-10pt]
\hline
& & \\[-10pt]
Einasto~\cite{Einasto} & $\displaystyle \rho_{s}\exp\left[-\frac{2}{0.17}\left[\left(\frac{r}{r_{s}}\right)^{0.17}-1\right]\right]$ & 28.44 & 0.033 \\
& & \\[-10pt]
\hline
& & \\[-10pt]
Burkert~\cite{Burkert}  & $\displaystyle  \frac{\rho_{s}}{(1+r/r_{s})(1+ (r/r_{s})^{2})}$ & 12.7 &  0.712 \\
\end{tabular}}
\label{tab:profiles}
\end{center}
where, in order to fix the parameters $r_s$ and $\rho_s$ at their precise values, one imposes the constraints that $\rho_\odot$ (the value of the DM density at the location of the solar system) $=0.3\, {\rm GeV}/{\rm cm}^3$ and that the total DM mass contained in the Milky Way reproduces observations (see Ref.~\cite{CCHHKPRSS}). They differ most at the Galactic Center (GC): NFW is peaked as $r^{-1}$ while Burkert is constant in the inner 1 kpc. They are instead similar around the location of the solar system, due also to the $\rho_\odot$ constraint.
As long as a convergent determination of the actual DM profile is not reached, it is sensible to have at disposal the whole range of these possible choices when computing Dark Matter signals in the Milky Way. In other words, the ignorance on the actual DM profile constitutes a (currently) irreducible astrophysical uncertainty for the predicted fluxes.

\medskip

\noindent {\bf Charged particles (electrons, positrons, antiprotons).} The $e^-$, $e^+$ and $\bar p$ produced in any given point of the halo propagate immersed in the turbulent galactic magnetic field. The field consists of random inhomogeneities that act as scattering centers for charged particles, so that their journey can effectively be described as a diffusion process from an extended source (the DM halo) to some final given point (the location of the Earth, in the case of interest). 
The number density $n_f(\vec x,E)$ per unit energy $E$  of the cosmic ray species $f$ $(= e^+,e^-, \bar p$) in any given point $\vec x$ evolves according to a diffusion-loss equation~\cite{SalatiCargese} 
\beq
\label{eq:diffeq}
-\mathcal{K}(E) \cdot \nabla^2 n_f - \frac{\partial}{\partial E}\left( b(E,\vec x) \, n_f \right) + \frac{\partial}{\partial z}\left( {\rm sign}(z)\, V_{\rm conv} \, n_f \right) = Q(E,\vec x) -2h\, \delta(z)\, \Gamma \, n_f\,.
\eeq
The first term accounts for diffusion, with a coefficient conventionally parameterized as $\mathcal{K}(E)=\mathcal{K}_0 (E/\GeV)^\delta$. The second term describes energy losses: the coefficient $b$ is position-dependent since the intensity of the magnetic field (which determines losses due to synchrotron radiation) and the distribution of the photon field (which determines losses due to inverse Compton scattering) vary across the galactic halo.
The third term deals with convection while the last term accounts for nuclear spallations, that occur with rate $\Gamma$ in the disk of thickness $h \simeq 100$ pc.
The source, DM annihilations, is denoted by $Q$.
The different processes described above have a different importance depending on the particle species: the journey of electrons and positrons is primarily affected  by synchrotron radiation and inverse Compton energy losses, while for antiprotons these losses are negligible and convection and spallation dominate.

Eq.~(\ref{eq:diffeq}) is usually solved numerically in a diffusive region with the shape of a solid flat cylinder that sandwiches the galactic plane, with height $2L$ in the $z$ direction and radius $R=20\,{\rm kpc}$ in the $r$ direction. The location of the solar system corresponds to $\vec x_\odot  = (r_{\odot}, z_{\odot}) = (8.33\, {\rm kpc}, 0)$.
Boundary conditions are imposed such that the number density $n_f$ vanishes on the surface of the cylinder, outside of which the charged cosmic rays freely propagate and escape. The values of the propagation parameters $\delta$, $K_0$, $V_{\rm conv}$ and $L$ are deduced from a variety of (ordinary) cosmic ray data and modelizations.
It is customary to adopt the following sets, denoted with MIN, MED and MAX because they are found to minimize or maximize the final fluxes
\begin{center}
\begin{tabular}{c|cc|ccc|c}
 & \multicolumn{2}{c|}{Electrons or positrons} & \multicolumn{3}{c|}{Antiprotons}  \\
Model  & $\delta$ & $\mathcal{K}_0$ [kpc$^2$/Myr] & $\delta$ & $\mathcal{K}_0$ [kpc$^2$/Myr] & $V_{\rm conv}$ [km/s] & $L$ [kpc]  \\
\hline
MIN  & 0.55 & 0.00595 & 0.85 &  0.0016 & 13.5 & 1 \\
MED & 0.70 & 0.0112 & 0.70 &  0.0112 & 12 & 4  \\
MAX  & 0.46 & 0.0765 &  0.46 &  0.0765 & 5 & 15
\end{tabular}
\end{center}
As long as independent measurements do not allow to pin down more precisely the values of these parameters, the scatter among such different sets constitute an additional astrophysical uncertainty on the predicted DM fluxes, this time due to the propagation process.

The solution of Eq.~(\ref{eq:diffeq}) allows to compute the phenomenological quantity in which we are interested: the flux of 
cosmic rays received at Earth $d\Phi_{f}/dE = v_{f} \, n_f /4 \pi$ (where $v_f$ is the velocity of species $f$, equal to $c$ for $e^\pm$ but possibly different for mildly-relativistic $\bar p$). It turns out that, both for $e^\pm$ and for $\bar p$, the flux can be conveniently expressed as a convolution of the spectra at the interaction point with some universal functions that encapsulate the astrophysics of the `production and propagation' process.  More precisely, for $e^\pm$ one has
\beq
\label{eq:positronsflux}
\frac{d\Phi_{e^\pm}}{dE}(E,\vec x_\odot) =
\frac{v_{e^\pm}}{4\pi \, b(E,\vec x_\odot)}
\displaystyle \frac12 \left(\frac{\rho_\odot}{M_\chi}\right)^2 \langle \sigma v \rangle \int_E^{M_\chi} dE_{\rm s} \, \frac{d{ N}_{e^\pm}}{dE}(E_{\rm s}) \, {I}(E,E_{\rm s},\vec x_\odot),
\eeq
where $dN_{e^\pm}/dE$ are the spectra at the annihilation point and ${I}(E,E_{\rm s},\vec x_\odot)$ are (generalized) halo functions which are independent of the particle physics model: there is such a function for each choice of DM distribution profile and choice of $e^\pm$ propagation parameters. We are following here the formalism discussed in Ref.~\cite{CCHHKPRSS}, which allows in particular to take into account the spatial dependence of the energy loss coefficient $b$ for $e^\pm$ discussed above. We refer to Ref.~\cite{CCHHKPRSS} for all details, including an explicit form of $b$ and of the $I$ functions, and further references.
Similarly, for $\bar p$ one has
\beq
\frac{d\Phi_{\bar p}}{dE}(E,\vec x_\odot) = \frac{v_{\bar p}}{4\pi}
 \displaystyle  \left(\frac{\rho_\odot}{M_\chi}\right)^2 R(E) \, \frac{1}{2} \langle \sigma v\rangle \frac{d{ N}_{\bar p}}{dE} .
\label{eq:fluxpbar}
\eeq
where it is now the function $R(E)$ which contains the astrophysics: again, there is such a function for each choice of DM distribution profile and the choice of $\bar p$ propagation parameters.

\bigskip

\noindent {\bf Neutral particles (photons, neutrinos).} Neutral messengers produced by DM annihilation in any given point of the DM halo travel along a straight line to the Earth. Since absorption in the Galaxy is negligible, the flux from a given direction is the result of the contribution from all the Dark Matter intervening along the line of sight.
The integrated flux of gamma rays or neutrinos over a region $\Delta \Omega$, corresponding e.g.\ to the window of observation or the resolution of the telescope, is given by
\beq
\label{gammafluxI}
\frac{d \Phi_{\gamma,\nu}}{dE}(E) =  \frac{r_\odot}{4\pi} \displaystyle \frac{1}{2} \left(\frac{\rho_\odot}{M_\chi}\right)^2 \bar J \, \Delta \Omega \, \langle \sigma v\rangle \frac{d{N}_{\gamma,\nu}}{dE}, \quad {\rm with}\ \bar J =\frac{1}{\Delta \Omega} \int_{\Delta \Omega} \int_{\rm l.o.s.} \frac{ds}{r_\odot} \left(\frac{\rho(r(s,\theta))}{\rho_\odot}\right)^2,
\eeq
where $dN_{\gamma,\nu}/dE$ denotes as usual the spectra at the annihilation point and the average $J$ factor contains the integral along the line of sight (l.o.s.). Here the coordinate $r$, centered  on the GC, reads $r(s,\theta)=(r_\odot^2+s^2-2\,r_\odot\,s\cos\theta)^{1/2}$: $s$ runs along the l.o.s. and $\theta$ is the aperture angle between the direction of the l.o.s. and the axis connecting the Earth to the GC. For a fixed window $\Delta \Omega$, the value of $\bar J$ can span orders of magnitude depending on the choice of the DM profile, especially if the window is small and close to the region where the profiles differ most, i.e. the GC. We refer to Ref.~\cite{CCHHKPRSS} for some explicit values of $\bar J$ in selected windows.

\subsection{Results}

\begin{figure}[!tb]
   \includegraphics[scale=0.85]{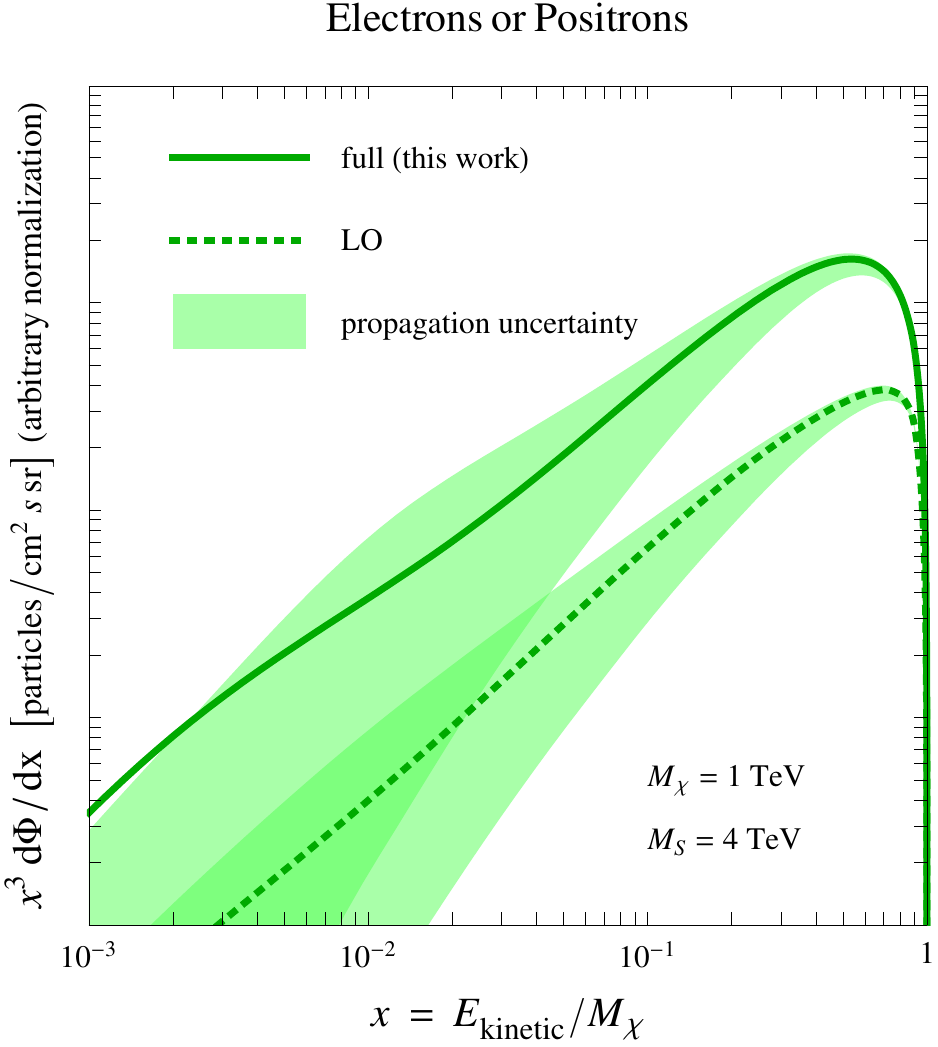}
   \quad
   \includegraphics[scale=0.85]{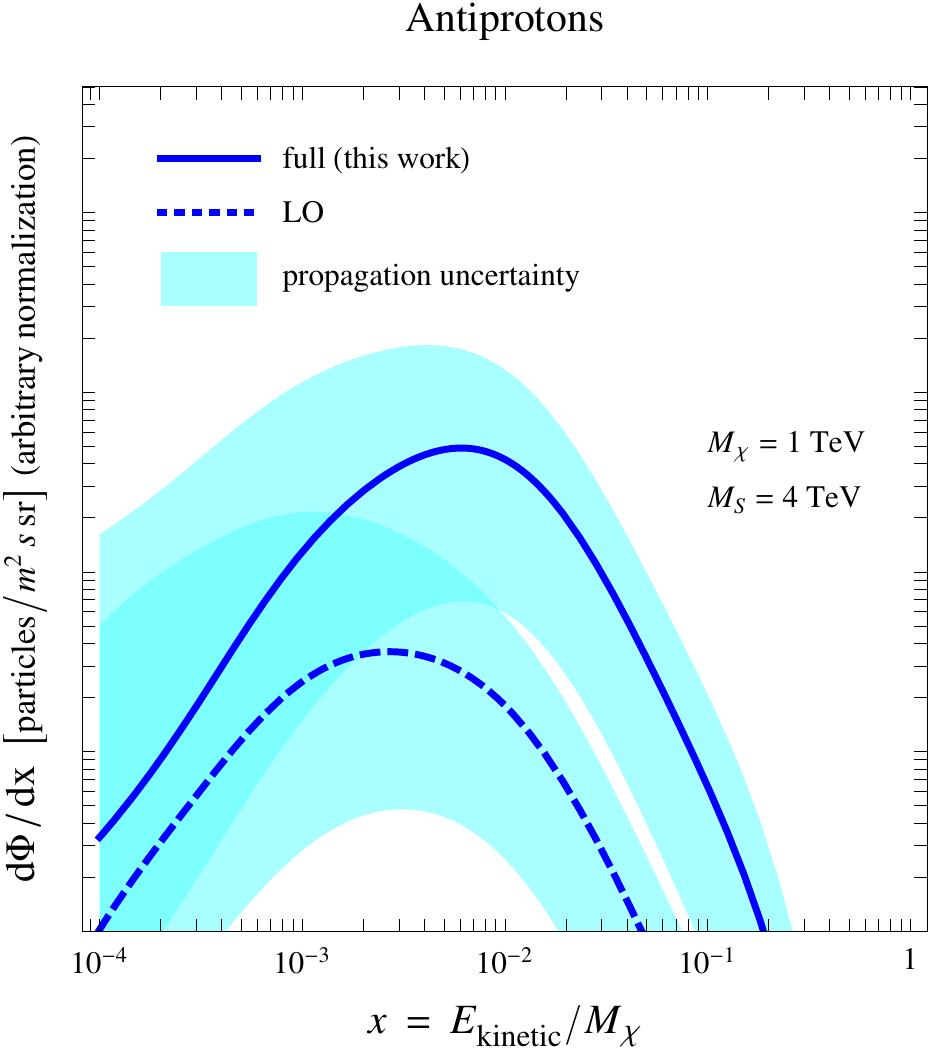}
 \caption{\emph{Fluxes of electrons or positrons \emph{(left panel)} and of antiprotons \emph{(right panel)}
after propagation in the galactic halo.}}
 \label{fig:propagated}
\end{figure}

In Figure~\ref{fig:propagated} we show the fluxes of charged cosmic rays for the
representative choice of model parameters already used in Section~\ref{sec:spectra}
(namely, $M_\chi = 1$ TeV and $M_S = 4$ TeV). 
We do not adopt here a specific value for the  annihilation cross section $\langle\sigma v\rangle$, which obviously
enters as a normalizing constant in Eqs.~(\ref{eq:positronsflux})
and~(\ref{eq:fluxpbar}): in the context of the toy model that we are considering,
its value is very small, if one requires the cosmological relic abundance of this DM
candidate to be fixed by the thermal freeze-out mechanism. This is not surprising
and it is actually the typical case for Bino-like Dark Matter in supersymmetry.
Notice that if one adopts the normalization with the 2-body annihilations, as done in 
Sect.~\ref{sec:spectra}, the energy spectra and the cross section appearing in the expressions
for the fluxes should be replaced by the analogous quantities in Eqs.~(\ref{eq:norm}) and (\ref{eq:set}).
What we are more interested in is verifying that the enhancement of the
fluxes remains significant on the phenomenological observables (the spectra after
propagation) and distinct from the normalization issues due to the propagation itself.
This is indeed the case, as Figure~\ref{fig:propagated} shows. In these plots, the
solid lines represent the fluxes computed with `NFW' as the choice of DM profile and
with `MED' propagation parameters. The shaded bands show the variance of this
prediction that one obtains by making other choices. The bands span quite a large
area since both the DM profile and the propagation parameters are variated
simultaneously.

The fluxes of $e^\pm$ and $\bar p$ are about one order of magnitude higher than
those computed in the LO approximation, consistently with what
expected from input fluxes, and 
emerge quite clearly from the uncertainty bands, especially for electrons
or positrons at high energies.
 Small differences in the shapes of the spectra, that were marginally visible in the
input spectra, see Figure~\ref{fig:spectra}, are essentially washed out for electrons
and positrons (since the energy losses tend to smooth any spectral feature) but
remain somewhat discernible in the antiproton spectrum (since propagation does not
significantly reshuffle energies for this species).

\medskip
In the case of neutral particles (gamma rays and neutrinos), the fluxes `at
detection' are easily computed with the use of Eq.~(\ref{gammafluxI}). They simply
correspond to a re-normalization of the input fluxes, for a given choice of the
observational window and the DM profile (which fixes the $\bar J$ factor), so that we do
not plot them explicitly. Any peculiar spectral feature possibly introduced by
contributions beyond the LO approximation, e.g.~the high-energy bump discussed above in the
gamma ray spectrum, would of course be conserved.

\section{Summary and conclusions}
\label{sec:conclusions}

We have investigated the relevance of the EW corrections in theories where the  
cross section for DM annihilation into 2-body final states is suppressed.
A Majorana DM annihilating into two light SM fermions is one such case.
We have worked for simplicity with a  model where the DM  is a Majorana fermion  of mass $M_\chi$ 
and a SM singlet, which annihilates into SM fermions  through the exchange of a heavy scalar doublet
of mass $M_S$, and carried out an expansion in $1/r\equiv (M_\chi/M_S)^2\ll 1$.

Let us summarize our main results:
\begin{itemize}
\item at the lowest order ($1/r$ in the amplitude) the radiation of EW gauge bosons is not able to remove
 the helicity suppression and the process stays in the $p$-wave
 (see Eq.~(\ref{estimate2body}) for an estimate and Eq.~(\ref{bpwave}) for the precise result);
\item an efficient removal of the suppression, opening up a potentially large $s$-wave, is achieved by including EW radiation
at the next-to-leading order ($1/r^2$ in the amplitude), which comes from both FSR and VIB diagrams
 (see Eq.~(\ref{estimate3body}) for an estimate and Eqs.~(\ref{eq:SwaveFinale})-(\ref{eq:PwaveFinale}) for the precise result);
\item the resulting energy spectra of stable particles, in the annihilation region, get substantially enhanced by this effect by 
factors $\mathcal{O}(10-100)$ (see Figure~\ref{fig:ratios}).
\end{itemize}
Furthermore, such an effect does not get spoiled by galactic propagation and crucially affects the predictions 
for fluxes to be measured at Earth (see Figure~\ref{fig:propagated}).

We have also interpreted our findings in the language of effective field theory and pointed out that
the effect of opening up the $s$-wave is missed by dimension-six operators and only catched
by higher-dimensional operators.
This is an example where the naive dimensional power counting fails to assess the relative 
importance of the operators in the expansion, as far as the EW radiation is concerned.

Our results have a wider generality than the specific model we have considered.
Reliable computations of energy spectra of stable particles and predictions for their fluxes at Earth -- 
the key observable for DM indirect searches -- cannot prescind from including the effects
of EW radiation.

\section*{Acknowledgments}

We are grateful to Stephen Mrenna, Torbj\"orn Sj\"ostrand,  and Peter Skands for useful
correspondence about {\sc Pythia}, and to Paolo Torrielli for  helpful advices on numerical
issues.
The work of ADS is supported by the Swiss National Science Foundation under
contract 200021-125237.
The work of AU is supported by CICYT-FEDER-FPA$2008$-$01430$.

\appendix

\section{The Dirac case}
\label{app:dirac}

We list here the analogous results of Sections \ref{sec:2body} and \ref{sec:3body}
for the case of Dirac Dark Matter.

The 2-body annihilation cross section into a pair of massless left-handed fermions  is
\begin{equation}
 v\sigma =a+b\,v^{2}+\mathcal{O}(v^4)\ ,
\end{equation}
where
\be
a=\frac{|y_L|^4}{32\pi (1+r)^2M_{\chi}^2}\, ,\qquad
b=\frac{|y_L|^4(r^2-3r-1)}{96\pi M_{\chi}^2(1+r)^4}\, .
\ee
For the process in Eq.~(\ref{eq:3Body}), the exchanged diagrams of Figure~\ref{fig:3Body} are
absent and the amplitude  can be written as
\begin{equation}
i\mathcal{M}\cdot\epsilon^{*}=\frac{ig|y_L|^2(1-2s_W^2)}{4c_W}\left[
\mathcal{M}_{A}+
\mathcal{M}_{B}
+\mathcal{M}_{C}
\right]\, ,
\end{equation}
and by using the Fierz transformation (\ref{eq:Fierz2}) the three terms analog of Eqs.~(\ref{eq:A})-(\ref{eq:C}) become
\begin{eqnarray}
\mathcal{M}_A&=&
\frac{\bar{u}_{f}\slashed{\epsilon}^*(k)(\slashed{p}_1+\slashed{k})P_R\gamma^{\mu}v_{f}}{2p_1\cdot k+m_Z^2}\cdot \left( \frac{D_{22}}{2}\bar{v}_\chi \gamma_{\mu}u_\chi
+\frac{D_{22}}{2}\bar{v}_\chi \gamma_{\mu}\;\gamma_5\;u_\chi
\right)\,, \\
\mathcal{M}_B&=&
(-1)[\bar{u}_{f}P_R\gamma^{\mu}v_{f}]\left[
(k_1-k_2-p_1+p_2)\cdot\epsilon^*(k)\;\bar{v}_\chi P_L\gamma_{\mu}u_\chi \;D_{11}\; D_{22}
\right] \\
\mathcal{M}_C&=&-\frac{\bar{u}_{f}P_R\gamma^{\mu}(\slashed{p}_2+\slashed{k})\slashed{\epsilon}^*(k)v_{f}}{2p_2\cdot k+m_Z^2}
\cdot \left( \frac{D_{11}}{2}\bar{v}_\chi \gamma_{\mu}u_\chi
+\frac{D_{11}}{2}\bar{v}_\chi \gamma_{\mu}\gamma_5 u_\chi
\right)\,.
\end{eqnarray}
Then the calculation of the cross sections proceeds as described in Section \ref{subsec:xsec} and
we choose the parametrization in terms of $\rho_s$ and $\rho_p$ as in Eq.~(\ref{eq:TotalCrossSection3Body}).
In particular for the partially-inclusive cross section in the large $M_S$ limit we find, neglecting terms vanishing in the $m_Z\to 0$ limit
\bea
\label{rissDirac}
{d\rho_s\over d x_2}&=&\frac{1}{6r^2}\left[(1-x_2)\left[-6+\frac{6(x_2+3)}{r}+\frac{(4x_2^3-26x_2^2-26x_2-30)}{r^2}\right]\right.\nonumber\\
&&\left.+\frac{1+x_2^2}{1-x_2}\left(3-\frac{6}{r}+\frac{9}{r^2}\right)\ln\frac{\overline{x}_+}{\overline{x}_-}\right]\;+\mathcal{O}(r^{-5})\, ,\\
{d\rho_p\over d x_2}&=&\frac{v^2}{36 r^2}\left[
(1-x_2)\left[-12+\frac{3(7x_2+35)}{r}+\frac{(18x_2^3-137x_2^2-227x_2-392)}{r^2}
\right]\right.\nonumber\\&&
\left.
+\frac{6(1+x_2^2)}{1-x_2}\left(1-\frac{7}{r}+\frac{21}{r^2}\right)\ln\frac{\overline{x}_+}{\overline{x}_-}\right]
\;+\mathcal{O}(r^{-5})\, .
\eea
For completeness, we also evaluate the expressions for the fully-inclusive $s$-wave and $p$-wave contributions
\bea\label{eq:fullyDirac1}
{\rho_s }&=& \frac{1}{60 r^2}\left[
30\left(1-\frac{2}{r}+\frac{3}{r^2}\right)\ln\frac{2M_{\chi}}{m_Z}\left(2\ln\frac{2M_{\chi}}{m_Z}-3\right)
\right.\nonumber\\&&\left.+5(\pi^2-15)-\frac{10(\pi^2-11)}{r}+\frac{3(5\pi^2-34)}{r^2}\right]\; ,\\
{\rho_p }&=& \frac{v^2}{360 r^2}\left[
30\ln\frac{2M_{\chi}}{m_Z}\left[4\left(1-\frac{7}{r}+\frac{21}{r^2}\right)\ln\frac{2M_{\chi}}{m_Z}-3\left(1-\frac{12}{r}+\frac{39}{r^2}
\right)\right]
\right.\nonumber\\&&
\left.+{5\over 2}(33-4\pi^2)+\frac{5(14\pi^2-155)}{r}+\frac{42(42-5\pi^2)}{r^2}\right]\; .
\eea

\section{3-body cross section in the $v\to 0$ limit}
\label{app:vzero}

We report here the results for the cross section of the 3-body process $\chi\chi\to f\bar f Z$,
in the limit $v\to 0$, therefore retaining only the part of the process proceeding through the $s$-wave.
We do not expand in powers of $1/r$, so the following results are valid for any value of $r\geq 1$.
The cross section is parametrized as in Eq.~(\ref{eq:TotalCrossSection3Body})
\begin{equation}
v\sigma|_{v\to 0}=\frac{\alpha_W|y_L|^4(1-2s_W^2)^2}{64\pi^2 c_W^2 M_\chi^2}\rho_s^{(v=0)}\,.
\end{equation}
The  partially-inclusive contribution to the differential cross section is
\bea\nonumber
{d\rho_s^{(v=0)}\over dx_2}&=& \frac{\frac{m_Z^2}{4 M_{\chi }^2}+x_2}{\left(r+x_2\right)^2}\left[  \frac{
   \frac{m_Z^2}{M_{\chi }^2}+2x_2 \left(r+1\right)+r^2-1}
   {4 \left(r+x_2\right)}
   \log\left[\frac{r+x_2-\bar y}{r+x_2+\bar y}\right]
   \right.\\
&&-\left.
   {\bar y\over 2}\frac{
   \frac{m_Z^2}{M_{\chi }^2}-2 x_2
   \left(r+x_2-1\right)-r^2-1}
   { \frac{m_Z^2}{M_{\chi }^2}+2\, x_2
   \left(r+1\right)+r^2-1}\right]\, ,
\eea
with $ \bar y=\sqrt{(1-x_2)^2-\frac{m_Z^2}{M_{\chi}^2}} $.
The fully-inclusive cross section is obtained by integrating the previous expressions over the kinematical
domain in Eq.~(\ref{eq:phasespace}). Neglecting terms vanishing in the limit $m_Z\to 0$ we find
\bea
\rho_s^{(v=0)}&=& \frac{1}{4r(1+r)}[A(r) r^3+B(r) r^2+C(r) r+D(r)]\, ,
\eea
where
\bea
A(r) &=& {\rm Li}_2\left(\frac{r-1}{2r}\right)-{\rm Li}_2\left(\frac{r+1}{2r}\right)+\ln(r+1)\ln\frac{r}{r^2-1}\nonumber\\&&+
\ln(r-1)\ln\frac{(r+1)^2}{r}+(\ln2-2)\ln\frac{r+1}{r-1}\; ,\\
B(r) &=& 2\left[
{\rm Li}_2\left(\frac{r-1}{2r}\right)-{\rm Li}_2\left(\frac{r+1}{2r}\right)+
\left(\ln\frac{r}{r+1}+\ln2-\frac{1}{4}\right)\ln\frac{r+1}{r-1}+2
\right]\; ,\\
C(r) &=& {\rm Li}_2\left(\frac{r-1}{2r}\right)-{\rm Li}_2\left(\frac{r+1}{2r}\right)+\left(
\ln\frac{r}{r+1}+\ln2+2\right)\ln\frac{r+1}{r-1}+3\; ,\\
D(r) &=& \frac{1}{2}\ln\frac{r+1}{r-1}\; ,
\eea
being ${\rm Li}_2(z)\equiv \sum_{k=1}^{\infty}z^k/k^2$ the usual dilogarithm.

\bibliographystyle{JHEP}


\end{document}